\documentclass[pdflatex,sn-basic,icol]{sn-jnl}

\jyear{2021}%

\usepackage{widetext}
\newcommand{\figref}[1]{Fig.~\ref{#1}}
\newcommand{\Figref}[1]{Figure~\ref{#1}}
\newcommand{\secref}[1]{Sec.~\ref{sec:#1}}

\newcommand{\appref}[1]{Appendix~\ref{append:#1}}
\renewcommand{\Re}{\operatorname{Re}}

\renewcommand{\eqref}[1]{Eq.~(\ref{eq:#1})}
\newcommand{\Eqref}[1]{Equation~(\ref{eq:#1})}
\renewcommand{\vec}[1]{\mathbf{#1}}
\newcommand{\mat}[1]{\mathbf{#1}}
\newcommand{\tr}{\operatorname{tr}}
\newcommand{\rev}{\color{black}}
\begin{document}

\title{Trace formulation for photonic inverse design with incoherent sources}


\author[1]{\fnm{Wenjie}~\sur{Yao}}\email{jayyao@mit.edu}

\author[2]{\fnm{Francesc}~\sur{Verdugo}}\email{fverdugo@cimne.upc.edu}

\author[3,4]{\fnm{Rasmus~E.}~\sur{Christiansen}}\email{raelch@mek.dtu.dk}
\author*[5]{\fnm{Steven~G.}~\sur{Johnson}}\email{stevenj@mit.edu}

\affil*[1]{\orgdiv{Department of Electrical Engineering and Computer Science}, \orgname{Massachusetts Institute of Technology}, \orgaddress{\city{Cambridge}, \state{MA}, \country{USA}}}

\affil[2]{\orgdiv{CIMNE}, \orgname{Centre Internacional de M\`{e}todes Num\`{e}rics a l'Enginyeria}, \orgaddress{\city{Castelldefels}, \country{Spain}}}

\affil[3]{\orgdiv{NanoPhoton--Center for Nanophotonics}, \orgname{Technical University of Denmark}, \orgaddress{\city{Kgs.~Lyngby}, \country{Denmark}}}
\affil[4]{\orgdiv{Department of Mechanical Engineering}, \orgname{Technical University of Denmark}, \orgaddress{\city{Kgs.~Lyngby}, \country{Denmark}}}

\affil[5]{\orgdiv{Department of Mathematics}, \orgname{Massachusetts Institute of Technology}, \orgaddress{\city{Cambridge}, \state{MA}, \country{USA}}}


\abstract{Spatially incoherent light sources, such as spontaneously emitting atoms, naively require Maxwell's equations to be solved many times to obtain the total emission, which becomes computationally intractable in conjunction with large-scale optimization (inverse design).  We present a trace formulation of incoherent emission that can be efficiently combined with inverse design, even for topology optimization over thousands of design degrees of freedom. Our formulation includes previous reciprocity-based approaches, limited to a few output channels (e.g. normal emission), as special cases, but generalizes to a continuum of emission directions by exploiting the low-rank structure of emission problems. We present several examples of incoherent-emission topology optimization, including tailoring the geometry of fluorescent particles, a periodically emitting surface, and a structure emitting into a waveguide mode, as well as discussing future applications to problems such as Raman sensing and cathodoluminescence.}




\maketitle


\section{Introduction}

Incoherent emission (light emission from random current sources) arises in many problems in optics: spontaneous emission (fluorescence)~\citep{Milonni1976SemiclassicalAQ,IncoherentSE,TraceFormula3}, thermal emission in both far~\citep{farthermal} and near~\citep{nearthermal,TraceFormula2} fields, scintillation~\citep{ICL,Charles}, Casimir and van~der~Waals forces~\citep{Casimir3}, Raman scattering in fluid suspensions~\citep{RamanReview}, incoherent incident waves~\citep{IncoherentImaging} (which can be transformed to random sources via the equivalence principle~\citep{EquivalencePrinciple}), and even scattering from surface roughness via a Born approximation~\citep{JohnsonPo05}.   However, accurate modeling of such spatially random sources can pose severe computational challenges, because a direct approach would involve averaging the results of many simulations over an ensemble of sources~\citep{MonteCarlo1,MonteCarlo3, Bao2019}; the statistics (correlation functions) of the \emph{sources} are known, but the difficulty is converting this into statistics (e.g. average power) of the resulting \emph{fields}.  In the cases of fluorescence~\citep{TraceFormula3}, near-field thermal radiation~\citep{TraceFormula2}, and Casimir forces~\citep{Casimir3}, for example, tractable methods for arbitrary geometries were only obtained recently.  This challenge is compounded when one wishes to perform \emph{inverse design}~\citep{InverseDesign}---large-scale optimization of emission over many geometric parameters, perhaps even over ``every pixel'' of a design region via topology optimization (TopOpt)~\citep{TOReview}---because one must then repeat the computation 10s--1000s of times as the design evolves, e.g.~to maximize spontaneous emission~\citep{SEinDielectricTheo,MaterialLossQ, RasmusCavity, LDOSOpt} or Raman emission~\citep{RamanTO} from a single molecule, much less a distribution of sources.

In this paper, we present a unified framework for inverse design of incoherent emission, combining a \emph{trace formulation} adapted from recent work~\citep{TraceFormula2,TraceFormula3,TraceFormula4} (\secref{TraceFormula}) with a new algorithm to \emph{simultaneously} optimize the geometry and evolve to an accurate estimate of the average emission/trace (\secref{resdom}).  We apply this framework to perform density-based TopOpt~\citep{TOReview} on several example problems in two dimensions: fluorescence from an optimized nanoparticle (\secref{averagedSE}), enhanced emission from a corrugated surface analogous to a light-emitting diode~\citep{LED} (\secref{periodic}), and optimized emission into a waveguide (\secref{waveguide}).  In each case, the emission is not from a single molecule, but the average power produced by an ensemble of incoherent emitters at every point in some material.  We show that this emission can be computed by a small number of ``eigen-sources'' of a Hermitian operator, which can be determined by a Rayleigh-quotient optimization~\citep{rayleigh} that is \emph{combined} with the inverse-design (geometric) optimization.   In the special case of emission into a small number $K$ of channels, such as $K$~far-field directions,  $K$~waveguide modes, or $K$~points in space, we show that a simple algebraic manipulation transforms the problem into $K$ simulations (\secref{fewcha})---this unifies and generalizes known results based on Kirchhoff's law of thermal radiation~\citep{Kirchhoff,Greffet2018} or (more generally) reciprocity~\citep{Charles,Janssen2010}  for computing emission into a single planewave direction---but our alternative approach (\secref{resdom}) yields a small number of solves even for $K\to\infty$.  The other well known special case is that of a single emitter location with a random orientation, which reduces to the local density of states (LDOS) via three Maxwell solves~\citep{Milonni1976SemiclassicalAQ,OskooiJo13-sources}, and this appears as another low-rank special case in our formulation~(\secref{fewincha}).  We believe that this computational framework will enable many future developments in the computational design of complex optical devices involving a wide variety of incoherent processes (\secref{conclusions}).


Density-based TopOpt has attracted increasing interest over the last few decades because of its ability to reveal surprising high-efficiency designs by optimizing over thousands or even millions of design degrees of freedom~\citep{TOReview}. It parameterizes a structure by an artificial ``density'' $\rho(\vec{x}) \in [0,1]$ at \emph{every point} (or every ``pixel'') in a design region, which is typically passed through smoothing and threshold steps to yield a physical ``binary'' design consisting of one of two materials at every point.
We apply a damped-diffusion filter~\citep{HelmholtzFilter}, which regularizes the problem by setting a minimum lengthscale on the design.  (Additional manufacturing constraints can be imposed by well-known techniques~\citep{FoundryConstraint}, but in the present work we focus on the fundamental algorithms and not on experimental realization.)  Once a scalar objective function (to be optimized) is defined, such as the emitted power (e.g.~the new formulation in this paper), its derivatives (sensitivities) with respect to all the design parameters can be efficiently computed with a single additional simulation via adjoint methods~\citep{InverseDesign,DesignSensitivity}.  Given the objective function and its derivatives, a variety of large-scale optimization algorithms are available; we use the CCSA/MMA method~\citep{svanberg2002class}.  We employ a recent free/open-source finite-element method (FEM) package, Gridap.jl~\citep{gridap}, in the Julia language~\citep{Julia}, which allows us to efficiently code highly customized FEM-based trace formulations in a high-level language, with the construction of the adjoint problem aided by automatic-differentiation (AD) tools~\citep{ForwardDiff,Zygote}.

\section{Trace Formulation} \label{sec:TraceFormula}

In this section, we first review the formulation of the frequency-domain Maxwell equations as a linear equation, discretized for numerical computation, with physical quantities like power as quadratic forms.  Then we show how the ensemble average of such an expression over a distribution of random current sources can be rewritten as a deterministic trace formula.  Finally, we explain how such a trace formula can be evaluated efficiently in the context of photonics optimization, both in the ``easy'' cases of coupling to a small number of output/input channels as well as in the more general cases of a continuum of outputs.

\subsection{Wave sources and quadratic outputs}

In the frequency domain, the linear Maxwell equations for the electric field $\vec{E}$ in response to a time-harmonic current source at a frequency $\omega$ are~\citep{FEMBook}
\begin{equation}
    \left[\nabla\times\frac{1}{\mu}\nabla\times-\left(\frac{\omega}{c}\right)^2\varepsilon\right]\vec{E}=\vec{f} \, ,\label{eq:me1}
\end{equation}
where $\varepsilon(\vec{x},\omega)$ is the relative electric permittivity, $\mu$ is the relative magnetic permeability ($\mu \approx 1$ for most materials at optical and infrared wavelengths, so we assume $\mu=1$ throughout this work), $c$ is the speed of light in vacuum, and $\vec{f}=\mathrm{i}\omega \vec{J}$ is a current-source term. 

Numerically, one discretizes the problem (e.g. using finite elements~\citep{FEMBook}) into a linear equation:
\begin{equation}
    \mat{A}\vec{u}=\vec{b} \, ,\label{eq:me2}
\end{equation}
where $\mat{A}$ is a matrix representing the Maxwell operator on the left-hand of \eqref{me1}, $\vec{u}$ is a vector representing the discretized electric (and/or magnetic) field, and $\vec{b}$ is a vector representing the discretized source term.   In the following, it is algebraically convenient to work with such a discretized (finite-dimensional) form, to avoid cumbersome infinite-dimensional linear algebra, but one could straightforwardly translate to the latter context as well~\citep{PCBook}.

Most physical quantities $P$ of interest in photonics---such as power (via the Poynting flux), energy density, and force (via the Maxwell stress tensor)---can be expressed as quadratic functions of the electromagnetic fields $\vec{u}$.   Since these are real-valued quantities, they correspond in particular to Hermitian quadratic forms
\begin{equation}
P = \vec{u}^\dagger \mat{O} \vec{u} \, ,
\label{eq:quadratic-P}
\end{equation}
where $\dagger$ denotes the conjugate transpose (adjoint) and $\mat{O}=\mat{O}^\dagger$ is a Hermitian matrix/operator.    In this paper, we are mainly concerned with computing emitted \emph{power} $P$, which is constrained by the outgoing boundary conditions to be non-negative, in which case $\mat{O}$ must furthermore be a \emph{positive semi-definite} Hermitian matrix  (i.e.,~non-negative eigenvalues) in the subspace of permissible $\vec{u}$, a property that will be useful in \secref{resdom}.

\subsection{Trace formula for random sources}

Now, consider the case where one has an ensemble
of random current sources $\vec{b}$ drawn from some statistical distribution with zero mean and a known correlation function (e.g.~a known mean-square current at each point if they are spatially uncorrelated).   In this case, we wish to compute the ensemble average, denoted by $\langle \cdots \rangle$, of our quadratic form \eqref{quadratic-P}:
\begin{equation}
    \langle P \rangle = \left\langle\vec{u}^\dagger\mat{O}\vec{u}\right\rangle = \left\langle\vec{b}^\dagger\mat{A}^{-\dagger}\mat{O}\mat{A}^{-1}\vec{b}\right\rangle \, ,\label{eq:obj1}
\end{equation}
where $\mat{A}^{-\dagger}$ denotes $(\mat{A}^{-1})^\dagger = (\mat{A}^\dagger)^{-1}$.  Note that only $\vec{b}$ is random in the right-hand expression.

Naively, this average could be computed by a brute-force method in which one explicitly solves the Maxwell equations ($\vec{u} = \vec{A}^{-1}\vec{b}$) for many possible sources $\vec{b}$ and then integrates over the distribution, perhaps by a Monte-Carlo (random-sampling) method.   That approach is possible, and has been accomplished e.g.~for evaluating thermal radiation~\citep{MonteCarlo1,MonteCarlo3}, but is computationally expensive.  Worse, such a direct approach quickly becomes prohibitive in the context of inverse design, where the averaging must be repeated for many geometries over the course of solving an optimization problem using an iterative algorithm. 

Instead, we adapt ``trace formula'' techniques that have been developed for similar problems in thermal radiation~\citep{TraceFormula2} and spontaneous emission~\citep{TraceFormula3}, where one must compute the average effect of many random current sources distributed throughout a volume. The basic trick (as reviewed in yet another related setting in \citep{TraceFormula4}) is to write the scalar $\langle P \rangle$ as a $1 \times 1$ ``matrix'' trace, and then employ the cyclic-shift property~\citep{LinearAlgebra} to group the $\vec{b}$ terms together:
\begin{widetext}
\begin{equation}
    \langle P \rangle = \left\langle\vec{b}^\dagger\mat{A}^{-\dagger}\mat{O}\mat{A}^{-1}\vec{b}\right\rangle=\tr\left\langle\vec{b}^\dagger\mat{A}^{-\dagger}\mat{O}\mat{A}^{-1}\vec{b}\right\rangle=\tr\left[\mat{A}^{-\dagger}\mat{O}\mat{A}^{-1}\langle\vec{b}\vec{b}^\dagger\rangle\right] \, . \label{eq:obj2}
\end{equation}
\end{widetext}
Here, the ensemble average is now confined to the
$\langle\vec{b}\vec{b}^\dagger\rangle$ term, which is just the correlation matrix~$\mat{B}$~\citep{Correlation} of the currents; such a matrix is positive semi-definite, so it can be factorized~\citep{TrefethenBook} (for convenience below) as
\begin{equation}
    \langle\vec{b}\vec{b}^\dagger\rangle=\mat{B}=\mat{D}\mat{D}^\dagger\, ,\label{eq:DD}
\end{equation}
for some known matrix $\mat{D}$.  Further information about constructing the matrix $\mat{B}$ or its factorization $\mat{D}$ is given in \appref{correlation}. (For the case of finite-element discretizations, we show that $\mat{B}$ is a sparse matrix that is straightforward to assemble and $\mat{D}$ is, for example, a sparse Cholesky factor~\citep{DavisBook}.)  Algebraically, expressing our results in terms of $\mat{D}$ below leads to convenient Hermitian matrices, but we show in \appref{factorization-free} that the final algorithms can easily employ $\mat{B}$ directly to avoid the computational cost of an explicit factorization.   In the simple case where random currents are \emph{spatially uncorrelated}, which holds for spontaneous emission and thermal emission in local materials~\citep{FDT}, $\mat{B}$ and $\mat{D}$ are conceptually \emph{diagonal} linear operators whose diagonal entries are the mean-square and root-mean-square currents, respectively, at each point in space.  Whether this leads to a strictly diagonal \emph{matrix} depends on the discretization scheme as explained in \appref{correlation}.   For instance, in the case of thermal and quantum fluctuations, the mean-square currents are given by the fluctuation--dissipation theorem (FDT)~\citep{FDT}, while for spontaneous emission one can use the FDT with a ``negative temperature'' determined by the population inversion~\citep{PickCe15,NegaTemp}.

Inserting \eqref{DD} into \eqref{obj2}, we obtain our objective as the trace of a deterministic Hermitian matrix $\mat{H}$ (which is positive-semidefinite if $\mat{O}$ is, as for power), given by:
\begin{equation}
    \langle P \rangle =\tr\underbrace{\left[(\mat{A}^{-1}\mat{D})^\dagger\mat{O}(\mat{A}^{-1}\mat{D})\right]}_\mat{H}\, .\label{eq:traceobj}
\end{equation}
The challenge now is to efficiently compute such a matrix trace.
Evaluating a trace is easy once the matrix elements are known---it is the sum of the diagonal entries---but the difficulty in \eqref{traceobj} is the computation of $\mat{A}^{-1}\mat{D}$.
Recall that the $N \times N$ matrix $\mat{A}$ is a discretized Maxwell operator where $N$ is the number of grid points (or basis functions), a huge matrix (especially in 3D).
There are fast methods to solve for $\mat{A}^{-1} (\mat{D} \vec{v})$ for any \emph{single} right-hand side $\vec{v}$, typically because the matrix $\mat{A}$ is \emph{sparse} (mostly zero) as in finite-element methods~\citep{FEMBook}, but computing the \emph{whole} matrix $\mat{A}^{-1}\mat{D}$ corresponds to solving $N$ right-hand sides.  Equivalently, computing explicit (dense) matrix inverses $\mat{A}^{-1}$ is typically prohibitively expensive (in both time and storage) for matrices arising in large physical systems~\citep{DavisBook}.   Fortunately, a large number of ``iterative'' algorithms have been proposed for \emph{estimating} matrix traces to any desired accuracy using relatively \emph{few} matrix--vector products~\citep{Hutchinson, FastEstimationTrace2017}, and what remains is to find a method well-suited to inverse design.

\subsection{Trace computation: Few output channels}\label{sec:fewcha}

In the important special cases where the desired output is the power in a small number ($K$) of discrete directions/channels/ports, or perhaps the intensity at a few points in space, we show in this section that the trace computation~\eqref{traceobj} simplifies to only $K$ scattering problems.   This fact is a generalization of earlier results commonly derived from electromagnetic reciprocity~\citep{Reciprocity}, such as the well-known Kirchhoff's law of thermal radiation (reciprocity of emission and absorption)~\citep{Kirchhoff} or analogous results for scintillation~\citep{Charles}.  More generally, this simplification arises whenever the matrix $\mat{O}$ in \eqref{quadratic-P} is \emph{low rank}.

For example, suppose that the objective function is the electric field intensity $\|\vec{E}(\vec{x}_1)\|^2$ at a single point $\vec{x}_1$ in space, which is the case for ``metalens'' optimization problems in which one is maximizing intensity at a focal spot~\citep{metalens}.   In matrix notation for a discretized problem, this quantity corresponds to 
\begin{equation}
   P = \|\vec{E}(\vec{x}_1)\|^2
   = \|\vec{e}_1^\dagger \vec{u}\|^2 = \vec{u}^\dagger \underbrace{\vec{e}_1\vec{e}_1^\dagger}_\mat{O} \vec{u} \; ,
\end{equation}
where $\vec{e}_1$ is the unit vector with a nonzero entry at the location (``grid point'') corresponding to $\vec{x}_1$.   We then have a rank-1~\citep{LinearAlgebra} matrix $\mat{O} = \vec{e}_1\vec{e}_1^\dagger$, and the trace~\eqref{traceobj} simplifies to $\langle P \rangle = \vec{v}_1^* \vec{v}_1 = \Vert \vec{v}_1 \Vert^2$ where
\begin{equation}
\vec{v}_1 = \mat{D}^\dagger \mat{A}^{-\dagger} \vec{e}_1 
\end{equation}
and $\mat{A}^{-\dagger} \vec{e}_1$ corresponds to solving a (conjugate-) \emph{transposed} Maxwell problem with a ``source'' $\vec{e}_1$ at the \emph{output} location, which is closely related to electromagnetic reciprocity~\citep{Reciprocity}.

Another important example where $\mat{O}$ is low-rank arises when the output $P$ is the power in one (or more) orthogonal ``wave channels''~\citep{OpticalWaveguide}, such as waveguide modes, planewave directions (e.g.~diffraction orders), or spherical waves.  In such cases the power in a given channel can be computed by squaring a \emph{mode-overlap integral} (e.g.~a Fourier component for planewaves) of the form $\Vert \vec{o}_1^\dagger \vec{u} \Vert^2$~\citep{OpticalWaveguide}.   Exactly as in the single-point case above, this corresponds to a rank-1 matrix $O = \vec{o}_1\vec{o}_1^\dagger$ and one must solve only a single ``reciprocal'' scattering problem to obtain the trace, where the ``source'' term is the (conjugated) \emph{output} mode $\vec{o}_1$.  This is precisely the situation in Kirchhoff's law, where in order to compute the average thermal radiation (emissivity) in a given direction, one solves a reciprocal problem for the absorption of an incident planewave in the opposite direction (the absorptivity)~\citep{Kirchhoff,Greffet2018,Janssen2010}.   A similar technique was recently applied to optimize the average power emitted in the normal direction from a scintillation device~\citep{Charles}.

More generally, such cases correspond to an output quadratic form $\mat{O}$ that takes a low-rank~\citep{LinearAlgebra} form:
\begin{equation}
    \mat{O} = \sum_{i=1}^K \vec{o}_i\vec{o}_i^\dagger\, ,\label{eq:Qchannel}
\end{equation}
where $K$ is the number of rank-1 terms $\vec{o}_i\vec{o}_i^\dagger$ (e.g. output channels/ports, output points, or other ``overlap integrals'').
Substituting \eqref{Qchannel} into \eqref{traceobj} and applying the cyclic-trace identity, we obtain:
\begin{eqnarray}
\langle P \rangle&=&\sum_{i=1}^K\tr\left[(\mat{A}^{-1}\mat{D})^\dagger \vec{o}_i\vec{o}_i^\dagger\mat{A}^{-1}\mat{D}\right]\nonumber \\
&=& \sum_{i=1}^K\vec{o}_i^\dagger\mat{A}^{-1}\mat{D}(\mat{A}^{-1}\mat{D})^\dagger\vec{o}_i\nonumber\\
&=& \sum_{i=1}^K \vec{v}_i^\dagger \vec{v}_i = \sum_{i=1}^K \Vert \vec{v}_i \Vert^2
\; ,\label{eq:objchannel}
\end{eqnarray}
where 
\begin{equation}
    \vec{v}_i=\mat{D}^\dagger\mat{A}^{-\dagger}\vec{o}_i \label{eq:vAq}
\end{equation}
corresponds to a single ``reciprocal'' Maxwell solve $\mat{A}^{-\dagger}\vec{o}_i = (A^{-T} \vec{o}_i^*)^*$ (a single scattering problem) for each~$i$.   (Electromagnetic reciprocity simply corresponds to the fact that $A^T = A$ for reciprocal materials~\citep{Reciprocity}.) Hence, the full trace---the average emission into $K$ channels---can be computed with only $K$ solves, and in many such cases $K=1$.

\subsection{Trace computation: Few input channels}\label{sec:fewincha}

One trivial special case in which the trace computation drastically simplifies is that of only a few sources or a few input channels, most famously in the case of the local density of states (LDOS): emission by a molecule at a \emph{single} location in space but with a random polarization~\citep{Milonni1976SemiclassicalAQ,OskooiJo13-sources}.  In the case of LDOS, this reduces the trace computation to three Maxwell solves, one per principal polarization direction, making the problem directly tractable for topology optimization~\citep{MaterialLossQ, RasmusCavity, LDOSOpt}.   More generally, this situation corresponds to the correlation matrix $\mat{B}$ being low-rank: if $\mat{B}$ is rank $K$, we can compute the trace in $K$ solves.

In particular, suppose that the currents $\vec{b}$ are of the form $\vec{b} = \sum_{i=1}^K \beta_i \vec{b}_i$ where the $\vec{b}_i$ are ``input channel'' basis functions (e.g. a point source with a particular orientation, or an equivalent-current source for a waveguide mode~\citep{OskooiJo13-sources}) and $\beta_i$ are uncorrelated random numbers with zero mean and unit mean-square.   Then the correlation matrix $\mat{B} = \langle \vec{b} \vec{b}^\dagger \rangle$  is simply the rank-$K$ matrix $\mat{B} = \sum_i \vec{b}_i \vec{b}_i^\dagger$.   In this case, the trace simplifies to:  
\begin{equation}
\tr\mat{H}= \sum_{i=1}^K \vec{u}_i^\dagger \mat{O}\vec{u}_i  \; ,\label{eq:inchannel}
\end{equation}
where computing $\vec{u}_i=\mat{A}^{-1}\vec{b}_i$ again requires only $K$ solves, one per source~$\vec{b}_i$.

\subsection{Trace computation: Many output channels}\label{sec:resdom}

In general, neither the matrix $\mat{O}$ nor the matrix $\mat{B}$ are low rank---for example, one may be interested in the total power radiated into a \emph{continuum} of angles above a surface, or some other infinite set of possible far-field distributions, from sources distributed over a continuous spatial region.   Fortunately, it turns out that there is another structure we can exploit: the Hermitian matrix $\mat{H} = (\mat{A}^{-1} \mat{D})^\dagger \mat{O} (\mat{A}^{-1} \mat{D})$ from \eqref{traceobj} is itself typically \emph{approximately low rank} (``numerically low rank''~\citep{LowRankApprox}) even if $\mat{O}$ is not: the trace, which equals the sum of the eigenvalues of~$\mat{H}$~\citep{LinearAlgebra}, is \emph{dominated by a few of $\mat{H}$'s largest eigenvalues}.   In this section, we first explain why that is the case, and then show how it can be exploited to efficiently estimate the trace during optimization.

There are two reasons to expect approximate low-rank structure of~$\mat{H}$ (which we illustrate with numerical examples in \secref{Application}).   First, on physical grounds, emission enhancement arises due to resonances (via the Purcell effect)~\citep{Purcell}, but in any finite volume there is some limit to the number of resonances that can interact strongly with emitters in a given bandwidth, related to an average density of states~\citep{SolarCell}. The traditional definition of resonant modes corresponds to poles of $\mat{A}^{-1}$ at complex resonant frequencies, which are (linear or nonlinear) eigenvalues $\omega$ satisfying $\det A(\omega)=0$ \citep{HMBook}; analogously, \eqref{traceobj} decomposes the total power into a sum of eigenvalues corresponding to ``resonant current'' sources which diagonalize~$\mat{H}$ at a given frequency. More explicitly, if $\mat{A}^{-1}\mat{D}$ can be accurately approximated by the action of~$K$ resonances of~$\mat{A}$ (a quasinormal mode expansion~\citep{Lalanne2018,Ge2014}), so that $\mat{A}^{-1}$ can be replaced by a rank-$K$ matrix, it follows that $\mat{H}$ is also approximately rank $\le K$ (since it is a product of rank-deficient matrices~\citep{LinearAlgebra}).  Moreover, geometric optimization to maximize the emitted power modifies the structure to further enhance one or more resonances~\citep{MaterialLossQ}, and we observe that this sometimes increases the concentration of the trace into a few eigenvalues of~$\mat{H}$; that is, optimized structures tend to be even lower rank.   Second, in a more general mathematical sense, the matrix $\mat{H}$ is built from \emph{off-diagonal blocks} of the Green's function matrix~$\mat{A}^{-1}$, connecting sources (at the the support of~$\mat{D}$) to emitted power at some other location (the support of~$\mat{O}$, e.g.~where the Poynting flux is computed), and off-diagonal blocks of Green's functions are known to be approximately low-rank~\citep{Hackbusch2015}.  This is closely related to fast methods for integral equations, such as the fast-multipole method and others~\citep{FMM}; essentially, far fields mostly depend on low-order spatial moments of the near fields/currents.

If $\tr \mat{H}$ is dominated by $K \ll N$ largest eigenvalues of the $N \times N$ matrix $\mat{H}$, then one merely needs a numerical algorithm to compute the $K$ \emph{extremal} (largest-magnitude) eigenvalues using only a sequence matrix--vector products $\mat{H}\vec{v}$ (corresponding to individual scattering problems).   Fortunately, there are many such algorithms, especially for Hermitian $\mat{H}$~\citep{Lanczos,LOBPCG}, and one can simply increase $K$ until the trace converges to any desired tolerance.   We argue here that methods based on Rayleigh-quotient maximization are particularly attractive for inverse design because they can be \emph{combined} with geometric/topology optimization.  The key fact is that one can express the sum of the largest $K$ eigenvalues as the maximum of a block Rayleigh quotient~\citep{rayleigh,JohnsonJo11,LOBPCG,Saad2011}, and for positive semidefinite $\mat{H}$ ($=$ positive semidefinite $\mat{O}$) this sum is a lower bound on the trace~\citep{Saad2011}:
\begin{widetext}
\begin{equation}
    \tr\mat{H}\geq \max_{\mat{V}\in\mathbb{C}^{N\times K}}\tr\left[(\mat{A}^{-1}\mat{D}\mat{V})^\dagger\mat{O}(\mat{A}^{-1}\mat{D}\mat{V})(\mat{V}^\dagger\mat{V})^{-1}\right]\, ,\label{eq:rayleighobj}
\end{equation}
\end{widetext}
where $\mat{V}$ represents any $K$-dimensional subspace basis, so that one is maximizing the trace over all possible subspaces.
This $\geq$ becomes equality for $N=K$, but in many problems (below) we find that $K < 10$ suffices for $< 1\%$ error in the trace (and, as expected from the arguments above, we find in \secref{averagedSE} that the required $K$ increases with the diameter of the emission region).  

Computationally, one can maximize the right-hand side of \eqref{rayleighobj} by some form of gradient ascent~\citep{rayleigh,LOBPCG}, each step of which only requires the evaluation of $\mat{A}^{-1}\mat{D}\mat{V}$ for a $N \times K$ matrix~$\mat{V}$.  That is to say, one only needs~$K$ Maxwell solves at each step (instead of~$N$ for the full matrix~$\mat{H}$), which vastly reduces the computational cost.

Moreover, this Rayleigh-quotient maximization formula is especially attractive in the context of inverse design, because it can be \emph{combined with the geometric optimization} itself. That is, instead of ``nesting'' the trace computation inside a larger geometric optimization procedure, we can simply add $\mat{V}$ to the geometry degrees of freedom and optimize over both $\mat{V}$ and the geometry \emph{simultaneously}. The full inverse-design problem with incoherent emission {\rev{can now be bounded by a \emph{single} optimization problem}}:
\begin{widetext}
\begin{equation}
    \langle P \rangle_\mathrm{optimum}\ge \max_{\mathrm{geometry},\mat{V}\in\mathbb{C}^{N\times K}}\tr\left[(\mat{A}^{-1}\mat{D}\mat{V})^\dagger\mat{O}(\mat{A}^{-1}\mat{D}\mat{V})(\mat{V}^\dagger\mat{V})^{-1}\right]\, ,\label{eq:rayleighopt}
\end{equation}
\end{widetext}
where the geometric  parameters (e.g.~material densities~\citep{TOReview} or level sets~\citep{LevelSet}) only affect~$\mat{A}$ and (perhaps)~$\mat{D}$, {\rev{and may be subject to some geometric and/or material constraints}}.  The gradient of the right-hand side with respect to the geometry can be computed efficiently with adjoint methods~\citep{InverseDesign,DesignSensitivity}, whereas the gradient with respect to $\mat{V}$ has a simple analytical formula~\citep{JohnsonJo11} (\appref{numform}), so a variety of gradient-based optimization algorithms~\citep{OptBook} can be applied to simultaneously evolve both $\mat{V}$ and the geometry. Furthermore, the Rayleigh quotient has the nice property that, since we are maximizing a lower bound on the full trace, the actual performance~$\langle P \rangle$ is guaranteed to be at least as good as the estimated performance at every optimization step.

\section{Topology-optimization formulation} \label{sec:Method}

In this section, we briefly review the density-based TopOpt formulation~\citep{TOReview} that we employ for our example applications in \secref{Application}.   The key idea of TopOpt is that an ``artificial density'' field, $\rho(\vec{x}) \in [0,1]$, is defined on a spatial ``design'' domain. This field is then filtered (to impose a non-strict minimum length-scale) and thresholded (to mostly ``binarize'' the geometry, resulting in a physically admissible geometry). The resulting smoothed and thresholded field is then used to control the spatial material distribution, constituting the structure under design. The design field, $\rho$, is discretized into a finite number of design degrees of freedom, which constitutes the design variables in the inverse design problem to be solved, e.g. \eqref{rayleighopt}, using a finite-element method (FEM) on a triangular mesh~\citep{gridap,FEMBook}, and the geometry is optimized using a well-known gradient-based algorithm that scales to high-dimensional problems with thousands or millions of degrees of freedom~\citep{svanberg2002class}.

Given a density $\rho(\vec{x}) \in [0,1]$, one should first regularize the optimization problem by setting a non-strict minimum lengthscale~$r_f$, as otherwise one may obtain arbitrarily fine features as the spatial resolution is increased.   This is achieved by convolving $\rho$ with a low-pass filter to obtain a smoothed density $\Tilde{\rho}$~\citep{TOReview}.  There are many possible filtering algorithms, but in an FEM setting (with complicated nonuniform meshes), it is convenient to perform the smoothing by solving a simple ``damped diffusion'' PDE, also called a Helmholtz filter~\citep{HelmholtzFilter}:
\begin{align}
    -r_f^2\nabla^2\Tilde{\rho}+\Tilde{\rho}&=\rho\, ,\nonumber\\
    \left. \frac{\partial \Tilde{\rho}}{\partial \vec{n}} \right\vert_{\partial\Omega_D} & =0 \, ,\label{eq:filter}
\end{align}
where $r_f$ is the lengthscale design parameter and $\vec{n}$ is the normal vector at the boundary~$\partial\Omega_D$ of the design domain~$\Omega_D$. This damped-diffusion filter essentially makes $\Tilde{\rho}$ a weighted average  of $\rho$ over a radius of roughly~$r_f$~\citep{HelmholtzFilter}.  (In addition to this filtering, it is possible to impose additional fabrication/lengthscale constraints, for example to comply with semiconductor-foundry design rules~\citep{FoundryConstraint}.)

Next, one employs a smooth threshold projection on the intermediate variable $\Tilde{\rho}$ to obtain a ``binarized'' density parameter $\Tilde{\Tilde{\rho}}$ that tends towards values of~$0$ or~$1$ almost everywhere~\citep{FilterThreshold}:
\begin{equation}
    \Tilde{\Tilde{\rho}} = \frac{\tanh(\beta\eta)+\tanh\left(\beta(\Tilde{\rho}-\eta)\right)}{\tanh(\beta\eta)+\tanh\left(\beta(1-\eta)\right)}\, ,\label{eq:threshold}
\end{equation}
where $\beta$ is a steepness parameter and $\eta = 0.5$ is the threshold.  During optimization, one begins with a small value of $\beta$ (allowing smoothly varying structures) and then gradually increases $\beta$ to progressively binarize the structure~\citep{RasmusTO}; here, we used $\beta=5,10,20,40,80$, similar to previous authors~\citep{RamanTO}.

Finally, one obtains a material, described by an electric relative permittivity (dielectric constant) $\varepsilon(\vec{r})$ in \eqref{me1}, given by:
\begin{equation}
    \varepsilon(\vec{r}) = \left[\varepsilon_1 +(\varepsilon_2-\varepsilon_1)\Tilde{\Tilde{\rho}}(\vec{r})\right]\left(1+\frac{\mathrm{i}}{2Q}\right) \, , \label{eq:toeps}
\end{equation}
where $\varepsilon_1$ is the background material (usually air, $\varepsilon_1=1$) and $\varepsilon_2$ is the design material (we use dielectric of $\varepsilon_2=12$ throughout this work).

\Eqref{toeps} includes an optional ``artificial loss'' term $\sim 1/ Q$, which effectively smooths out resonances to have quality factors $\le Q$ (fractional bandwidth $\ge 1/ Q$)~\citep{MaterialLossQ}.  Such an artificial loss is useful in single-$\omega$ emission optimization in order to set a minimum bandwidth of enhanced emission, rather than obtaining diverging enhancement over an arbitrarily narrow bandwidth as is possible with lossless dielectric materials~\citep{MaterialLossQ}.  Also, optimizing low-$Q$ resonances often leads to better-behaved optimization problems (less ``stiff'' problems with faster convergence), so during optimization we start with a low $Q = 5$ and geometrically increase it (to $Q = 1000$) as the optimization progresses~\citep{MaterialLossQ}

The details of the FEM discretization are described in \appref{numform}, but it is essentially a standard triangular mesh with first-order Lagrange elements~\citep{FEMBook} and perfectly matched layers (PMLs) for absorbing boundaries~\citep{PML}.   We discretized $\rho$ and $\{\tilde\rho,\tilde{\tilde\rho}\}$ with piecewise-constant (0th-order) and first-order elements, respectively.   During optimization, one must ultimately compute the sensitivity of the objective function (the trace from \secref{TraceFormula}) with respect to the degrees of freedom~$\rho$---for each step outlined above (smoothing, threshold, PDE solve, etcetera) we formulate a vector--Jacobian product following the adjoint method for sensitivity analysis~\citep{InverseDesign,DesignSensitivity} with some help from automation~\citep{ForwardDiff}, and then these are automatically composed (``backpropagated'') by an automatic-differentiation (AD) system~\citep{Zygote}.  In this way, the gradient with respect to all of the degrees of freedom ($\rho$ at every mesh element) can be computed with about the same cost as that of evaluating the objective function once~\citep{InverseDesign}.

\section{{\rev{Numerical examples}}} \label{sec:Application}

In this section, we present three example problems in 2D illustrating how our trace-optimization procedure works in practice for typical problems involving ensembles of spatially incoherent emitters. We start in \secref{averagedSE} with a general case where we are maximizing the total emitted power from many emitters distributed throughout a ``fluorescent'' dielectric material. Next, in \secref{periodic}, we study the enhanced emission from a corrugated surface, analogous to a light-emitting diode~\citep{LED}, showing how the trace formulation can be applied to a periodic structure with aperiodic emitters. Both of these examples are based on the general algorithm from \secref{resdom}, which can handle emission into a continuum of possible angles.   Finally, in \secref{waveguide} we apply the more specialized algorithm from \secref{fewcha} to optimizing emission from a fluorescent material into a single-mode waveguide. {\rev{Since Maxwell's equations are scale-invariant~\citep{PCBook}, the same optimal designs will be obtained for any wavelength $\lambda$ if the geometry (thickness and period) is scaled with $\lambda$ (for the same dielectric constants).}}

\begin{figure*}[tb]
\centering
\includegraphics[width=0.9\linewidth]{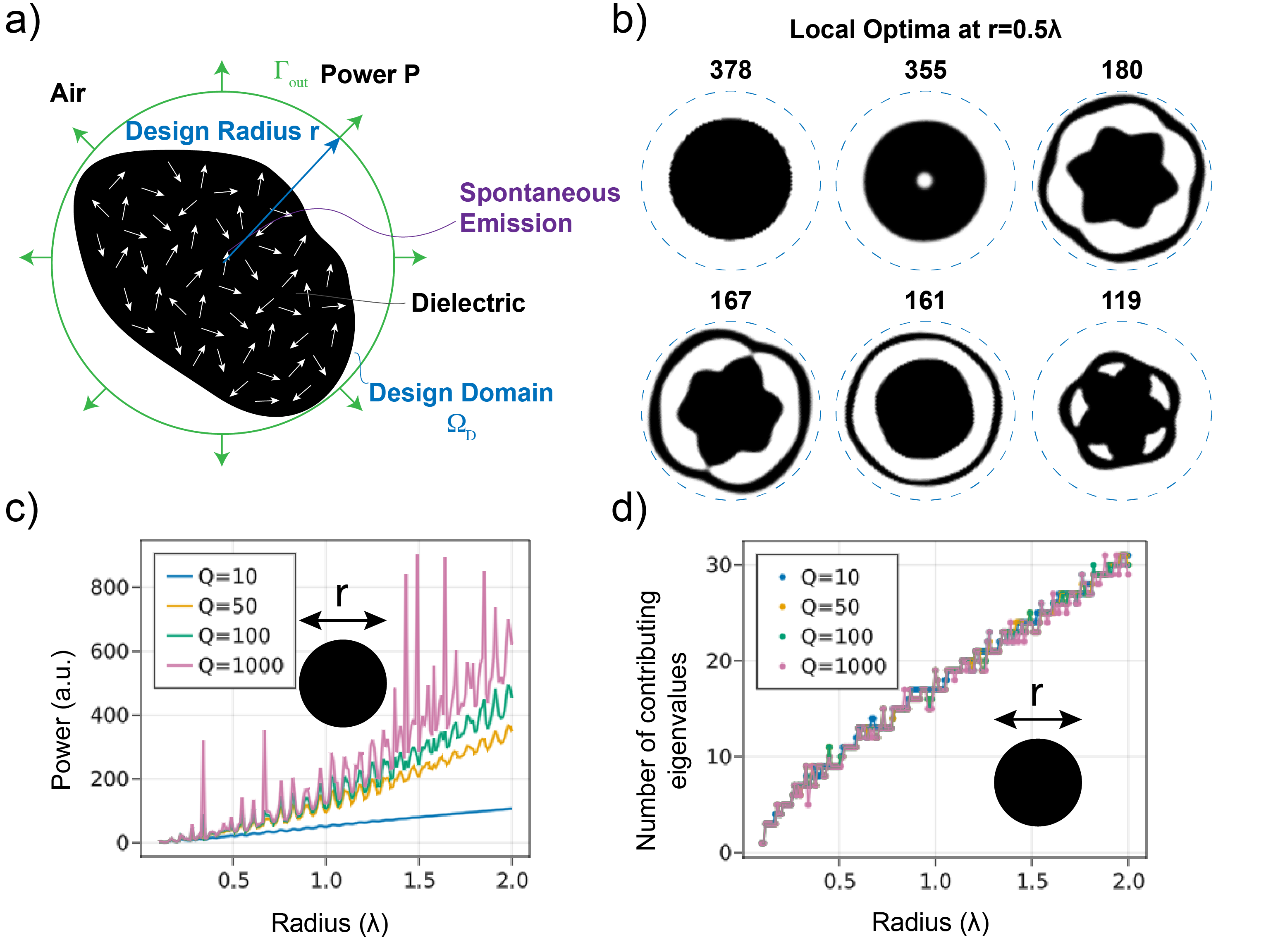}
\caption{(a) A 2D fluorescent particle (of dielectric $\varepsilon=12$) with a circular design domain of radius $r$. The emitters are distributed uniformly within the dielectric material. The total power $P$ radiated outwards in \emph{any} direction (integral of Poynting flux over $\Gamma_{\mathrm{out}}$) at a wavelength $\lambda$ is optimized. (b) Typical local optima found for design radius $r=0.5\lambda$ with filling-ratio $R_f=0.5$ and bandwidth quality factor $Q=1000$. The numbers above denote the optimized emitting (average) power in arbitrary unit. (c) Emitted power of a disk as a function of the disk radius $r$ for different bandwidth quality factors $Q$. (d) The number of eigenvalues that contribute 99\% of the trace as a function of the disk radius $r$ for different bandwidth quality factors $Q$. \label{fig1} }
\end{figure*}
\subsection{Fluorescent particle}\label{sec:averagedSE}

In this example, illustrated in \figref{fig1}a, we optimize the
shape/topology of a 2D fluorescent dielectric ($\varepsilon = 12$) particle constrained to have a given area lying within a circular design domain of radius~$r$, maximizing the total power $P$ radiated outwards in \emph{any} direction at a wavelength $\lambda$. The emitters are distributed uniformly within the dielectric material. Further computational details can be found in~\appref{averagedse}.

Because this is a non-convex optimization problem, topology optimization can converge to different local optima from different initial geometries~\citep{InverseDesign}. \Figref{fig1}b shows multiple local-optima geometries for a design radius $r=0.5\lambda$ with filling-ratio $R_f=0.5$ and bandwidth quality factor $Q=1000$ (artificial loss, from~\secref{Method}), obtained from different initial geometries (disks of different radii and/or $\varepsilon$). The numbers above the geometries denote the corresponding emitted (average) power $P$ in arbitrary units. In this particular case, after examining a large number of local optima (not shown), we found that the best local optimum is simply a circular disk with a particular radius. {\rev{The existence of many local optima with performance varying by factors of 2--5 is not unusual in wave problems~\citep{LDOSOpt,Diaz2010,BermelGh10}, and while various heuristic strategies have been proposed to avoid poor local minima~\citep{MutapcicBo09,Aage2017,BermelGh10,Schneider2019} beyond simply probing multiple random starting points, the only way to obtain rigorous guarantees is to derive theoretical upper bounds~\citep{millerbound,LDOSOpt} as discussed further in \secref{conclusions} (purely numerical global search can generally provide practical guarantees only for very low-dimensional Maxwell optimization~\citep{AzunreJe19}).}}

Whether the best optimum is a disk changes with the design-domain radius, and appears to depend on whether there is a nearby radius with a high-$Q$ resonance at the design~$\lambda$.  (In fact, for this particular case the locally optimal disk has an area slightly less than our upper bound, meaning that the area constraint is not active. In consequence, this particular disk remains a \emph{local} optimum even if the design domain is enlarged, and apparently remains a global optimum until the design domain is sufficiently enlarged to admit a stronger resonance. {\rev{Although the area constraint is not active at this particular local optimum, it is active at intermediate points during the optimization process, and there are many \emph{other} local optima that would also be found if the area constraint were not present.  Physically that emitted power can increased simply by adding more fluorescent material; correspondingly, without an area constraint we often find a local optimum in which the design region is almost entirely filled with dielectric.}}) In~\figref{fig1}c, we show how the average power radiated by a circular disk varies with radius $r/\lambda$, and clearly exhibits a series of sharp peaks correspond to radii which support high-$Q$ resonances at $\lambda$: the familiar whispering-gallery resonant modes~\citep{WGM2D}.

The key assumption of our algorithm in \secref{resdom} was that only a small number of eigenvalues would contribute to the trace, and this assumption clearly holds here.   In~\figref{fig1}d, we plot the number of eigenvalues that contribute 99\% of the trace as a function of the disk radius.
We can see that only a small number of eigenvalues is required to obtain a good estimate of the trace; we find similar results for other shapes.  Naively, one might expect that the number of contributing eigenvalues would scale with the area (or volume in 3d), corresponding to the number of resonances per unit bandwidth from the density of states (DOS)~\citep{SolarCell}.  However, we find that the scaling is nearly linear with the disk radius; the reason the simple DOS argument fails is that it doesn't take into account the variable loss (radiation) rates of the modes, which causes most of the resonances to contribute weakly even if the \emph{real} part of their frequency is close to the emission frequency.  In fact, we have found similar linear scaling of the number of contributing eigenvalues for many other shapes, including other locally optimized shapes, and it appears to be an interesting open theoretical question to prove (or disprove) asymptotic linear scaling.

\begin{figure*}[tb]
\centering
\includegraphics[width=0.9\linewidth]{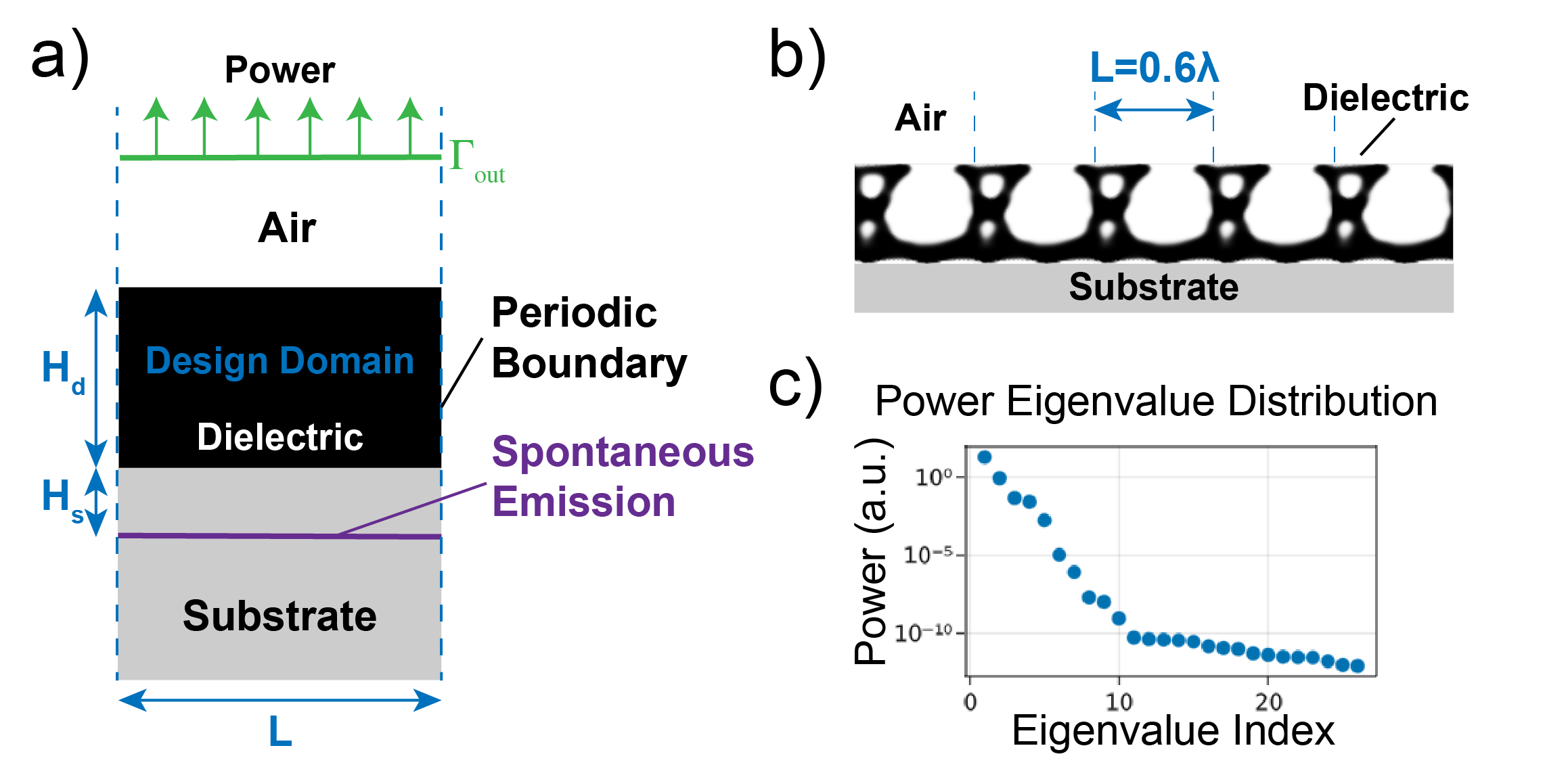}
\caption{(a) Unit cell of a 2D periodic emitting surface with period $L$. The design domain consists of dielectric material ($\varepsilon = 12$) in air with thickness $H_d=0.5\lambda$, the spontaneous-emission current sources are uniformly distributed on an horizontal line (purple line) inside a lower-index substrate ($\varepsilon=2.25$) a distance $H_s=0.1\lambda$ below the design domain. The objective is the total power emitted upwards, integrated over $\Gamma_{\mathrm{out}}$ (b) Optimized geometry with period~$L=0.6\lambda$. (c) The eigenvalue distribution of the average power for the optimized geometry.  \label{fig2} }
\end{figure*}
\subsection{Periodic emitting surface} \label{sec:periodic}

In this example, we enhance the emission from a thin "emitting layer" by optimizing a periodically patterned surface situated on top of the layer.---this is inspired by a light-emitting diode (LED) with a patterned surface above an active emitting layer, where it is well known that a periodic pattern can enhance emission via guided-mode resonances~\citep{LED, LEDReview}. As illustrated in \figref{fig2}a, the design domain consists of dielectric material ($\varepsilon = 12$) in air with a period $L$ and thickness $H_d=0.5\lambda$, the spontaneous-emission current sources are uniformly distributed on an horizontal line (``active layer'') inside a lower-index substrate ($\varepsilon=2.25$) a distance $H_s=0.1\lambda$ below the design domain. The objective, here, is the total power emitted  upwards, integrated over \emph{all angles} (i.e., the total Poynting flux) using the methods of \secref{resdom}.  (Emission purely into the normal direction could be optimized much more efficiently using the methods of \secref{fewcha}.)  Further computational details can be found in~\appref{periodic}.

Even though the dielectric structure is periodic (the design domain is a single unit cell of $\varepsilon$), the emitters are \emph{not} periodic---they are independent random currents at every point in the active layer.  Computationally, however, we can still reduce the simulation of \emph{non-periodic} sources in a periodic medium to a set of small \emph{unit-cell simulations}, using the ``array-scanning method''~\citep{sumk}.   An arbitrary aperiodic source current can be Fourier-decomposed into a superposition of Bloch-periodic sources ($\vec{J}_k(x+L)=e^{\mathrm{i}kL}\vec{J}_k(x)$), each of which can be simulated with a single unit cell and Bloch-periodic boundary conditions in $x$.   The total power is then simply obtained from an integral ($\int_{-\pi/L}^{\pi/L}\mathrm{d}k)$) over the Bloch wavevector $k$ in the Brillouin zone.   For incoherent aperiodic random sources, \emph{each} of these Bloch-periodic unit-cell calculations is an operator trace (over random currents in the unit cell only) computed by the methods of \secref{TraceFormula}.  (Unit-cell calculations for different $k$ values are completely independent and can be performed in parallel.) Further details of this formulation are described in \appref{periodic}.   (Moreover, the array-scanning method can be viewed as a special case of a reduction using symmetry: for any symmetry group, sources can be decomposed into a superposition of ``partner functions'' of the irreducible representations of the symmetry group~\citep{Inui2012group}, thus reducing the simulation domain even for asymmetrical random sources.)

The optimized structures for the design parameter~$H_d=0.5\lambda$,~$H_s=0.1\lambda$ is shown in \figref{fig2}b. Note that we have also optimized over the period $L$ (here, simply by repeating the optimization for different values of $L$) to find an optimized period~$L=0.6\lambda$. The eigenvalue distribution of the average power is given in \figref{fig2}c: Again, we observe that only the first few eigenvalues contribute significantly to the trace, as conjectured in \secref{resdom}.
\begin{figure*}[tb]
\centering
\includegraphics[width=0.99\linewidth]{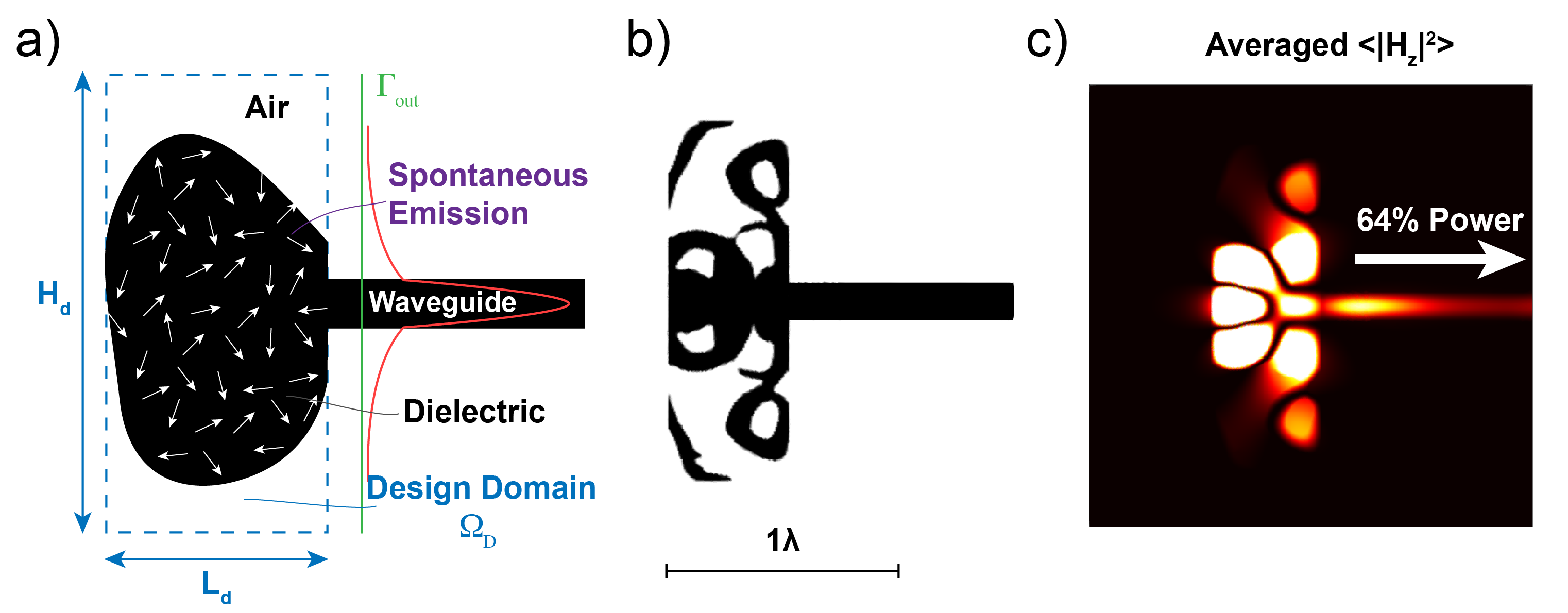}
\caption{(a) A 2D fluorescent dielectric ($\varepsilon=12$) medium coupling to a waveguide ($\varepsilon=12$, width~$\lambda/2\sqrt{12}$). The design domain is of height $H_d=1.5\lambda$ and width $L_d=0.5\lambda$, the power coupled into a single-mode dielectric waveguide (mode overlap integral at $\Gamma_\mathrm{out}$) is optimized. (b) Optimized shape with a filling-ratio $R_f=0.5$. (c) Averaged field intensity $\langle\vert H_z\vert^2\rangle$ distribution. About 64\% percent of the power is coupled into the waveguide mode. \label{fig3} }
\end{figure*}

\subsection{Emission into a waveguide} \label{sec:waveguide}

This example considers a fluorescent dielectric ($\varepsilon=12$) medium in air, similar to \secref{averagedSE}, but in this case we are maximizing the power coupled into a single-mode dielectric waveguide ($\varepsilon=12$, width~$\lambda/2\sqrt{12}$) rather than into radiation (\figref{fig3}a).   Since the output is a single channel ($O$ is rank~1), this allows us to apply the method of \secref{fewcha} to perform only a single ``reciprocal'' Maxwell solve per optimization step. Since the waveguide breaks the rotational symmetry of the problem, the optimum structure is now very different from a circular disk, and must somehow redirect light emitted \emph{anywhere} in the fluorescent material into the waveguide.   This task is made more difficult by the fact that we employ a design domain whose size is only $1.5\lambda \times 0.5\lambda$, so the optimization cannot simply surround the emitters with a multi-layer Bragg mirror to confine the radiation (as occurs when optimizing LDOS in a large design domain~\citep{MaterialLossQ, RasmusCavity}).
Further computational details can be found in~\appref{waveguide}.

\Figref{fig3}b shows the optimized geometry with a design domain of height $H_d=1.5\lambda$ and width $L_d=0.5\lambda$.  The material is constrained to fill at most half of the design domain (to illustrate that we can independently constrain the design region and the design volume); unlike for the disk optimum in \secref{averagedSE}, this area constraint was active at the optimum shown here. The corresponding averaged field intensity $\langle \vert H_z\vert^2 \rangle$ is displayed in \figref{fig3}c. We found that 64\% of the power is coupled into the desired waveguide mode. In comparison, only 4\% of the power is coupled to the waveguide mode for a trivial rectangular design where the the whole design domain is filled with $\varepsilon=12$ fluorescent material.

\section{Conclusion}
\label{sec:conclusions}

We presented a trace formulation and accompanying algorithms for topology optimization of incoherent emitters, which unify and generalize earlier work, and in particular provide the first tractable optimization algorithms for the challenging case of many random emitters and many output channels.   Looking forward, we believe that there are many potential applications of these ideas, as well as further algorithmic improvements and generalizations.

We are already preparing to use these techniques to optimize Raman sensing in fluid suspensions of many Raman molecules, in contrast to previous work that only considered a single molecule location~\citep{RamanTO,Raman3d}---it will help us to answer the interesting open question of the optimal spatial density of ``hot spots'' where light is concentrated to enhance Raman emission.  Another application is enhancing cathodoluminescence or other forms of scintillation detectors, which were previously optimized only for normal emission~\citep{Charles}.  In contrast to spontaneous emission, where the light is emitted by spatially uncorrelated \emph{point} sources, one can instead consider incoherent beams of light consisting of uncorrelated random \emph{planewave} amplitudes---this corresponds to spatially \emph{correlated} random currents~\citep{IncoherentImaging}, and we are investigating the resulting trace formulation to design metalenses for incoherent focusing.   Other applications include the study of radiation loss due to surface roughness, which can be modeled via random sources with a prescribed correlation function related to the manufacturing disorder and may naively require a large number of Maxwell solves~\citep{JohnsonPo05,KitaMi18,Payne1994}.   Nor is our approach limited to Maxwell's equations---it is applicable to any linear system where one wishes to optimize quadratic functions of random source terms.

Algorithmically, we are investigating ways to apply more sophisticated algorithms to the joint structure/trace optimization problem \eqref{rayleighopt}.   When solving the eigenproblem alone (maximizing over $\mat{V}$ to obtain extremal eigenvalues), it is well known that one can greatly improve upon straightforward gradient ascent by Krylov algorithms such as Arnoldi~\citep{TrefethenBook} or LOBPCG~\citep{LOBPCG}, and we would like to incorporate Krylov acceleration into to joint problem as well.  {\rev{Recent techniques to accelerate frequency-domain solves for multiple sparse inputs and outputs~\citep{lin2022full} may also be applicable to accelerate our trace optimization (since we have multiple sources in a sparse subset of the domain, and objective functions like the power only involve sparse outputs).}} Similar to the stochastic Lanczos algorithm~\citep{FastEstimationTrace2017}, one could further exploit the fact that we are computing the trace of a function $f(\mat{A})$ of the Maxwell operator $\mat{A}$ in order to relate the trace more efficiently to Krylov subspaces of $\mat{A}$.  More generally, there are other applications where one is maximizing $\tr f(\mat{A}(p),p)$ for some $f$ and some parameters $p$, and it seems similarly beneficial to combine the trace estimation with the parameter optimization in such problems.

{\rev{Theoretically, it is desirable to complement improved numerical optimizations with new rigorous upper bounds on incoherent emission.  Significant progress has already been made on bounding thermal-emission processes~\citep{MillerJo15,Rodriguez2020} as well as to absorption~\citep{Kuang2020,millerbound} (related to emission via reciprocity), and many of these techniques should be adaptable to other forms of random emission.}}

\backmatter

\section*{Funding}
This work was supported in part by the U.S. Army Research Office through the Institute for Soldier Nanotechnologies under award W911NF-13-D-0001, and by the PAPPA program of DARPA MTO under award HR0011-20-90016. F. Verdugo acknowledges support from the program Severo Ochoa Centre of Excellence (2019-2023) under the grant CEX2018-000797-S funded by MCIN/AEI/10.13039/501100011033. R. E. Christiansen acknowledges support from the Danish National Research Foundation (Grant No. DNRF147 - NanoPhoton).

\section*{Disclosure}
On behalf of all authors, the corresponding author states that there is no conflict of interest.
\begin{appendices}

\section*{Replication of results}
The code for \secref{Application} can be found at\\ \url{https://github.com/WenjieYao/TraceFormula}.
\section{Correlation matrix}
\label{append:correlation}

In this section, we show how to compute the correlation matrix $\mat{B}$ corresponding to random current sources $\vec{J}$ discretized in a finite-element basis.   One can express the frequency-domain Maxwell equations either in terms of the electric field $\vec{E}$, in which case the source term is proportional to $\vec{J}$, or in terms of the magnetic field $\vec{H}$, in which case the source term is proportional to $\nabla \times \vec{J}$~\citep{FEMBook}.  These two formulations lead to different $\mat{B}$ correlation matrices.

In particular, we consider the case where the currents $\vec{J}$ (at a frequency $\omega$) are spatially uncorrelated with a given correlation function:
\begin{equation}
    \left\langle\vec{J}(\vec{x})\vec{J}(\vec{x}^\prime)^\dagger\right\rangle = \mat{C}(\vec{x}) \delta(\vec{x}-\vec{x}^\prime)\, ,\label{Jcor}
\end{equation}
where $\mat{C}$ is a given $3\times 3$ Hermitian positive-semidefinite correlation matrix.   For example, in 2D with in-plane electric currents, as in the examples of \secref{Application}, one has
\begin{equation}
    \mat{C} = \begin{pmatrix} J_0^2 & & \\ & J_0^2 & \\ & & 0 \end{pmatrix},
    \label{eq:in-plane-C}
\end{equation}
where $J_0^2(\vec{x})$ is the mean-square current at~$\vec{x}$.  For isotropic random currents, $\mat{C}=J_0^2 \mat{I}$ where $\mat{I}$ is the identity matrix.

In a finite-element method, the source vector $\vec{b}$ is constructed by taking inner products of the source current with real vector-valued basis ``element'' functions $\hat{\vec{v}}_n$ (Nedelec elements in 3D, or $\hat{v}_n \hat{\vec{z}}$ with scalar Lagrange elements $\hat{v}_n$ in 2D for $z$-polarized fields)~\citep{FEMBook}.  That is, the components of $\vec{b}$ are
\begin{equation}
    b_n = \int \hat{\vec{v}}_n \cdot \mbox{(source current)} \, \mathrm{d}\Omega.
\end{equation}

For an electric-field formulation with a source current $\vec{J}$, we obtain the correlation function:
\begin{align}
    B_{mn} &= \langle b_m b_n^* \rangle\,, \nonumber\\
     &= \left\langle \iint \hat{\vec{v}}_m(\vec{x})^T \vec{J}(\vec{x}) \vec{J}(\vec{x}^\prime)^\dagger \hat{\vec{v}}_n(\vec{x}^\prime)  \, \mathrm{d}\Omega \mathrm{d}\Omega^\prime \right\rangle \,, \nonumber\\
     &=  \iint \hat{\vec{v}}_m(\vec{x})^T \left\langle \vec{J}(\vec{x}) \vec{J}(\vec{x}\prime)^\dagger\right\rangle \hat{\vec{v}}_n(\vec{x}^\prime) \, \mathrm{d}\Omega \mathrm{d}\Omega^\prime\,, \nonumber \\
     &=  \int \hat{\vec{v}}_m^T \mat{C} \hat{\vec{v}}_n \, \mathrm{d}\Omega \, .
\end{align}
For localized basis functions (as in a finite-element method), this results in an extremely sparse matrix $\mat{B}$---it is zero if $\hat{\vec{v}}_m$ and $\hat{\vec{v}}_n$ don't overlap, or in regions where the mean-square current $\mat{C}$ is zero.  (If $\mat{C}$ is the identity, $\mat{B}$ is equal to the Gram matrix of the basis.) Note also that, by construction, $\mat{B}$ is a Hermitian semidefinite matrix, so it has factorization $\mat{B} = \mat{D}\mat{D}^\dagger$, such as a Cholesky factorization~\citep{TrefethenBook}.

For a magnetic-field formulation, $\vec{J}$ is replaced by $\nabla\times \vec{J}$ above, but we can simply integrate by parts~\citep{PCBook} to move the $\nabla\times {}$ curl operation to act on the basis functions, yielding:
\begin{equation}
    B_{mn} = \langle b_m b_n^* \rangle = \int (\nabla\times\hat{\vec{v}}_m)^T \mat{C} (\nabla\times \hat{\vec{v}}_n) \, \mathrm{d}\Omega \, .
    \label{eq:B-curl}
\end{equation}
Again, this yields a sparse Hermitian semidefinite matrix $\mat{B}$.

In the 2D examples of \secref{Application}, we employed a magnetic-field formulation with an out-of-plane magnetic field $\vec{H} = H_z \hat{\vec{z}}$ and corresponding basis functions $\hat{v}_n \hat{\vec{z}}$, along with in-plane current sources corresponding to \eqref{in-plane-C}.  In this case, \eqref{B-curl} simplifies to:
\begin{equation}
    B_{mn} =\int_\Omega J_0^2\left(\nabla \hat{v}_m\cdot\nabla \hat{v}_n\right)\mathrm{d}\Omega \, .\label{eq:Bmn2D}
\end{equation}

\section{Factorization-free trace formulation}
\label{append:factorization-free}

Although it is conceptually attractive to use a trace formulation \eqref{traceobj} in terms of the Hermitian matrix $\mat{H}$, this formulation required a factorization $\mat{B} = \mat{D}\mat{D}^\dagger$ of the correlation matrix $\mat{B}$.   Computationally, it is desirable to avoid this factorization, especially if the current distribution (and hence $\mat{B}$) depends on the geometric degrees of freedom ${\rho}$ (which would require us to differentiate through the matrix factorization in our adjoint calculation).  Instead, it is straightforward to reformulate our optimization problems \eqref{objchannel} and \eqref{rayleighopt} in terms of $\mat{B}$ alone using a change of variables.

For the few-output-channel case in \secref{fewcha}, one can simply start with \eqref{traceobj} and rewrite it as $\langle P \rangle = \tr [ \mat{A}^{-\dagger} \mat{O} \mat{A}^{-1} \mat{B} ]$, which for a low-rank $\mat{O}$ simplifies, similar to \eqref{objchannel}, to
\begin{equation}
    g(\rho) = \langle P \rangle  =\sum_{i=1}^K \vec{u}_i^\dagger\mat{B}\vec{u}_i\, ,\label{eq:append_fewcha_obj}
\end{equation}
where $\mat{A}^\dagger\vec{u}_i=\vec{o}_i$, and we have defined the parameter $\rho$ dependence (which can effect both $\mat{A}$ and $\mat{B}$) as a function $g(\rho)$ for use in the adjoint formulation of \appref{numform}.

For the many-channel case of \eqref{rayleighopt}, the key point is that we can choose $\mat{V}$ to be orthogonal to the nullspace $N(\mat{D})$ of $\mat{D}$, as any nullspace component would contribute nothing to the trace ($\mat{D}\mat{V}$ projects it to zero).  Equivalently, we can choose $\mat{V}=\mat{D}^\dagger W$ ($\perp N(D)$~\citep{LinearAlgebra}) for any $N\times K$ matrix $\mat{W}$, 
and this change of variables yields a new optimization problem:
\begin{widetext}
\begin{equation}
   g(\rho,\mat{W})=\tr [\left(\mat{A}^{-1}\mat{B}\mat{W}\right)^\dagger\mat{O}\underbrace{\left(\mat{A}^{-1}\mat{B}\mat{W}\right)}_{\mat{U}}(\mat{W}^\dagger\mat{B}\mat{W})^{-1} ]= \tr\left[\mat{U}^\dagger\mat{O}\mat{U}(\mat{W}^\dagger\mat{B}\mat{W})^{-1}\right]
   \, ,\label{eq:append_tracemany}
\end{equation}
\end{widetext}
where again we have defined the function $g(\rho,\mat{W})$  for the parameter and $\mat{W}$ dependence, along with $\mat{U} = \mat{A}^{-1} \mat{BW}$, for use in the adjoint formulation of \appref{numform}.

\section{Numerical formulation}
\label{append:numform}

In this section, we provide details of the mathematical formulation and numerical implementation of the examples in \secref{Application}, including the adjoint analysis.

We employ the frequency-domain Maxwell equations for the magnetic field $\vec{H}$ arising from an electric current $\vec{J}$ with a dielectric function (relative permittivity) $\varepsilon$ and a relative magnetic permeability $\mu$:
\begin{equation}
    \left[\nabla\times\frac{1}{\varepsilon}\nabla\times-\left(\frac{\omega}{c}\right)^2\mu\right]\vec{H}(\vec{x})=\nabla\times\left[\frac{1}{\varepsilon}\vec{J}(\vec{x})\right] \, .\label{eq:append_meH}
\end{equation}
For 2D ($z$-invariant) problems, we chose in-plane currents $\vec{J}$, so that the resulting magnetic fields $\vec{H}=H_z\hat{\vec{z}}$ are polarized purely in the $z$ direction~\citep{PCBook}.  In this case \eqref{append_meH} simplifies to a scalar Helmholtz equation:
\begin{equation}
    \left[-\nabla\cdot\frac{1}{\varepsilon}\nabla-\left(\frac{\omega}{c}\right)^2\mu\right]H_z=\left(\nabla\times\left[\frac{1}{\varepsilon}\vec{J}(\vec{x})\right]\right)\cdot\hat{\vec{z}} \, .\label{eq:append_meHz}
\end{equation}
Note that, for the correlation functions in the previous discussion, we simplified the right-hand side by absorbing the $1/\varepsilon$ scaling into $\vec{J}$.

We employ perfectly matched layers (PMLs) for absorbing boundaries, with Dirichlet ($u=0$) boundary conditions behind the PML. The implementation of the ``stretched-coordinate'' PML is simply a replacement $\nabla\rightarrow\Lambda\nabla$ in \eqref{append_meHz}~\citep{PML,FEMBook}:
\begin{equation}
    \left[-\Lambda\nabla\cdot\frac{1}{\varepsilon}\Lambda\nabla-\left(\frac{\omega}{c}\right)^2\mu\right]H_z=\left(\nabla\times\left[\frac{1}{\varepsilon}\vec{J}(\vec{x})\right]\right)\cdot\hat{\vec{z}} \, .\label{eq:append_meHz_pml}
\end{equation}
where
\begin{equation}
    \Lambda = \begin{pmatrix} \frac{1}{1+\mathrm{i}{\sigma_x(\vec{x})}/{\omega}} & & \\ & \frac{1}{1+\mathrm{i}{\sigma_y(\vec{x})}/{\omega}} & \\ & & \frac{1}{1+\mathrm{i}{\sigma_z(\vec{x})}/{\omega}} \end{pmatrix}.
    \label{eq:append_Lamda}
\end{equation}
The PML conductivity $\sigma_\ell(\vec{x})$, $\ell=x,y,z$ function is used to gradually ``turn on'' the PML to compensate for discretization errors~\citep{PML}, and we use a quadratic profile $\sigma_\ell(\vec{x})=\sigma_0(x_{\mathrm{PML}}/d_{\mathrm{PML}})^2$ (where $x_{\mathrm{PML}} \in [0,d_{\mathrm{PML}}]$ is the distance inside the PML).

\subsection{Fluorescent particle}
\label{append:averagedse}
For the problem of \secref{averagedSE}, the governing equation is exactly \eqref{append_meHz_pml} with $\mu = 1$, whose weak form is~\citep{FEMBook}:
\begin{eqnarray}
    a(u,v)&=&b(v),\nonumber\\
    a(u,v)&=&\int_\Omega (\nabla \Lambda v\cdot\frac{1}{\varepsilon}\Lambda\nabla u-k_0^2 vu)\mathrm{d}\Omega\, ,\nonumber\\
    b(v)&=&\int_\Omega vf\mathrm{d}\Omega\, , \label{eq:weakformSE}
\end{eqnarray}
where $k_0=\omega/c$ is the free-space wave number, $f=(\nabla\times\vec{J})\cdot\hat{\vec{z}}$ is the source term, and $\nabla\Lambda$ denotes the linear operator $\nabla\Lambda u = \nabla(\Lambda u)$. The matrix $\mat{A}$ and the source vector $\vec{b}$ for the discretized Maxwell equation \eqref{me2} are obtained by replacing $u$ and $v$ with the finite-element basis functions $\hat{u}_n$ and $\hat{v}_n$, using first-order Lagrange elements on a triangular mesh~\citep{FEMBook}.  The mesh was generated with Gmsh~\citep{gmsh}, corresponding to a spatial resolution of roughly $\lambda/40$ in the air and $\lambda/80$ in the design region.

Notice that in \eqref{append_tracemany}, only $\mat{U}$ (via $\mat{A}$) and $\mat{B}$ (describing emission only in the dielectric) depend on the design parameters $\rho$. {\rev{We have now the optimization problem as:}}

\begin{eqnarray}
   g(\rho,\mat{W})&=& \max_{\rho,\mat{W}}\tr\left[\mat{U}(\rho)^\dagger\mat{O}\mat{U}(\rho)(\mat{W}^\dagger\mat{B}(\rho)\mat{W})^{-1}\right]
   \, ,\nonumber\\
   \mat{U}(\rho) &=& \mat{A}(\rho)^{-1}\mat{B}(\rho)\mat{W}\,,\nonumber\\
   0&\leq&\rho\leq1\,,\nonumber\\
   \int \rho\mathrm{d}\Omega_d&<&\int R_f\mathrm{d}\Omega_d\,,\label{eq:append_fpobjective}
\end{eqnarray}
{\rev{where $R_f$ is the area filling-ratio.}}

Applying adjoint-method analysis~\citep{InverseDesign,DesignSensitivity}, we obtain the partial derivatives:
\begin{widetext}

\begin{multline}
    \frac{\partial g}{\partial \rho}=-\tr\left[\mat{U}^\dagger\mat{O}\mat{U}(\mat{W}^\dagger\mat{B}\mat{W})^{-1}(\mat{W}^\dagger\frac{\partial \mat{B}}{\partial p}\mat{W})(\mat{W}^\dagger\mat{B}\mat{W})^{-1}\right] \\ -2\Re\left\{\tr\left[\mat{Z}^\dagger\left(\frac{\partial \mat{A}}{\partial \rho}\mat{U}-\frac{\partial \mat{B}}{\partial \rho}\mat{W}\right)\right]\right\}\, ,\label{eq:dgdpSE}
\end{multline}
\end{widetext}
where $\mat{Z}$ is the result of an adjoint solve:
\begin{equation}
    \mat{A}^\dagger\mat{Z}=\mat{O}\mat{U}(\mat{W}^\dagger\mat{B}\mat{W})^{-1}\, . \label{eq:adjointZ}
\end{equation}
The partial derivative with respect to $\mat{W}$ is simply obtained via matrix~\citep{MatrixCookbook} CR~calculus~\citep{CRCalculus}:
\begin{widetext}

\begin{equation}
    \frac{\partial g}{\partial\mat{W}}=\left[\mat{I}-\mat{B}\mat{W}(\mat{W}^\dagger\mat{B}\mat{W})^{-1}\mat{W}^\dagger\right](\mat{A}^{-1}\mat{B})^\dagger\mat{O}\mat{U}(\mat{W}^\dagger\mat{B}\mat{W})^{-1}\, .\label{eq:dgdVSE}
\end{equation}
\end{widetext}

\begin{figure*}[tb]
\centering
\includegraphics[width=0.9\linewidth]{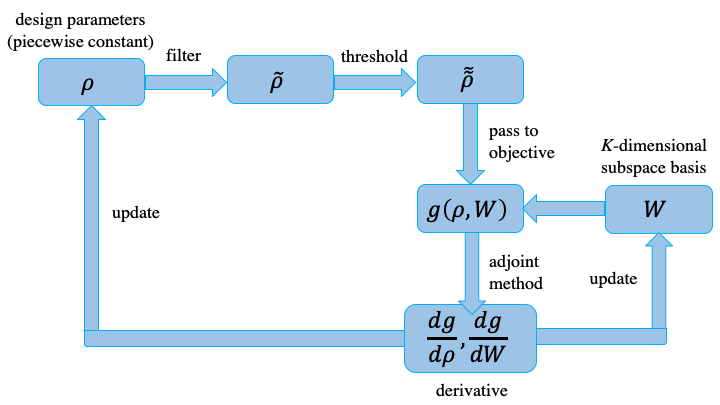}
\caption{Flowchart of the optimization steps for the fluorescent particle and periodic emitting surface examples.  \label{flowchart1} }
\end{figure*}

{\rev{We validated the derivatives from the adjoint method against finite differences at random points, and found that the relative error was only about $10^{-6}$ or less, which is not a problem for the CCSA algorithm when converging the optimum to only a few decimal places. 

The analysis workflow for this example is shown in \figref{flowchart1}. This CCSA update is implemented with NLopt in Julia~\citep{NLopt} for an increasing series of $\beta=5, 10, 20, 40, 80$. And for each $\beta$, the loop is terminated either a relative difference of $10^{-8}$ is achieved or the maximum iteration reaches 200. The design parameter $\rho$ is bounded from 0 to 1. }}

\subsection{Periodic emitting surface}
\label{append:periodic}

For the problem of \secref{periodic}, we simulate a single unit cell with Bloch-periodic boundary conditions in~$x$.  Since Gridap only supports periodic boundary conditions in its current version, we make a change of variables $H_z \rightarrow H_z e^{\mathrm{i}kx}$ so that $H_z$ is the periodic ``Bloch envelope'' function~\citep{PCBook}.  In comparison to \eqref{append_meHz} in \appref{averagedse}, this corresponds to the transformation
$\nabla\rightarrow \nabla + \mathrm{i}k\hat{\vec{x}}$~\citep{PCBook}:
\begin{equation}
   \left[-(\nabla+\mathrm{i}k\hat{\vec{x}})\cdot\frac{1}{\varepsilon}(\nabla+\mathrm{i}k\hat{\vec{x}}) -k_0^2\right] H_z = f\, ,\label{eq:append_me_periodic}
\end{equation}
with periodic boundaries in~$x$, whose weak form (including PML in $y$) can then be obtained via integration by parts:
\begin{widetext}
\begin{eqnarray}
    a(u,v)&=&b(v),\nonumber\\
    a(u,v)&=&\int_\Omega \left[\left(\nabla\Lambda-\mathrm{i}k\hat{\vec{x}}\right)v\cdot\frac{1}{\varepsilon}\cdot\left(\Lambda\nabla+\mathrm{i}k\hat{\vec{x}}\right)u-k_0^2 vu\right]\mathrm{d}\Omega,\nonumber\\
    b(v)&=&\int_\Omega vf\mathrm{d}\Omega\, . \label{eq:weakformP}
\end{eqnarray}
\end{widetext}
where $\Lambda$ is the diagonal PML ``stretching'' matrix~\eqref{append_Lamda}.

The objective (average power) is then constructed by a Brillouin-zone integration over the Bloch wavevector~$k$~\citep{sumk}:
\begin{equation}
    g(\rho)=\frac{L}{2\pi}\int_{-\pi/L}^{\pi/L}\tr\left[\left(\mat{A}_k^{-1}\mat{D}\right)^\dagger\mat{O}\left(\mat{A}_k^{-1}\mat{D}\right)\right]\mathrm{d}k\, ,\label{eq:sumgk1}
\end{equation}
where $L$ is the period of the unit cell and $\mat{A}_k$ is assembled using \eqref{weakformP}.  Since this integrand is a periodic function of~$k$, the integral can be approximated by a simple trapezoidal sum over equally spaced points $k$ with exponential accuracy~\citep{TrefethenExp}; we used $100$~$k$ points in order to resolve sharp resonances.

Commuting the integral and the trace in \eqref{sumgk1}, similarly to \appref{factorization-free} (noting that $\int \tr = \tr \int$), we obtain

\begin{eqnarray}
   g(\rho,\mat{W})&=&\max_{\rho,\mat{W}}\frac{L}{2\pi}\int_{-\pi/L}^{\pi/L}\tr\left[\mat{U}_k(\rho)^\dagger\mat{O}\mat{U}_k(\rho)(\mat{W}^\dagger\mat{B}(\rho)\mat{W})^{-1}\right]\mathrm{d}k\, ,
   \, ,\nonumber\\
   \mat{U}_k(\rho) &=& \mat{A}_k(\rho)^{-1}\mat{B}(\rho)\mat{W}\,,\nonumber\\
   0&\leq&\rho\leq1\,. \label{eq:sumgk2}
\end{eqnarray}

The adjoint analysis for \eqref{sumgk2} is almost the same as in \appref{averagedse}, except for the additional integration over~$k$. {\rev{Also, it shares the same analysis workflow as in \appref{averagedse}}}.

\subsection{Emission into a waveguide}
\label{append:waveguide}
\begin{figure*}[tb]
\centering
\includegraphics[width=0.9\linewidth]{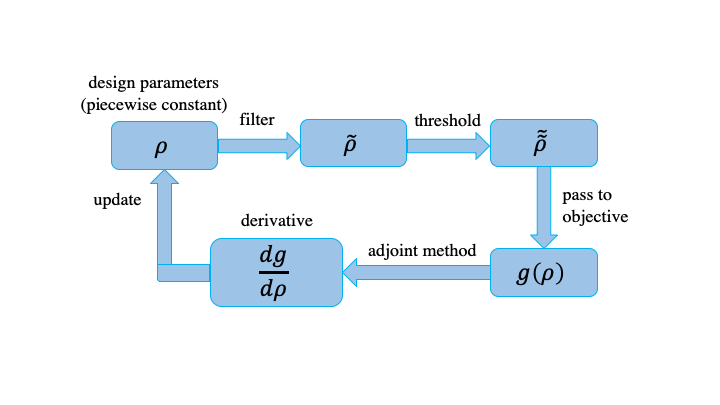}
\caption{Flowchart of the optimization steps for the emission into a waveguide example.  \label{flowchart2} }
\end{figure*}
For the problem of \secref{waveguide}, the governing equation and the weak form are identical to \appref{averagedse}.   The main difference is our objective function is now the power in a waveguide mode, computed via an overlap integral using mode orthogonality~\citep{OpticalWaveguide}, rather than a total Poynting flux.  Here, we briefly review how this overlap integral is implemented in the finite-element method.

For a propagating waveguide mode with electric and magnetic fields $\vec{e}_i$ and $\vec{h}_i$, the modal-expansion coefficient $\alpha_{i}$ of that mode for a total magnetic field $\vec{H}$ is given by the overlap integral~\citep{OpticalWaveguide}
\begin{equation}
    \alpha_{i}^{*} = \frac{\int \vec{e}_{i}\times\vec{H}^{*}\cdot \mathrm{d}\vec{S}}{\int \vec{e}_{i}\times\vec{h}_{i}^{*}\cdot \mathrm{d}\vec{S}}=\frac{\int e_{yi}H_z^*\mathrm{d}y}{\int e_{yi}h_{zi}^{*}\mathrm{d}y}\, ,\label{eq:modeoverlap}
\end{equation}
where we have assumed an $x$-oriented waveguide in 2D and an in-plane electric-field polarization.  The power carried by this mode is then simply $\vert\alpha_{i}\vert^2$.   In \secref{waveguide}, our objective is the power $\vert\alpha_{0}\vert^2$ in a single mode:
\begin{equation}
    \langle P\rangle= \vert\alpha_0\vert^2= \left\vert  \frac{1}{N_0} \int e_{y0}H_z^{*}\mathrm{d}y\right\vert^2\, ,\label{eq:gchan}
\end{equation}
where $N_0$ is the normalization (which can be omitted for optimization) from \eqref{modeoverlap}.  If $H_z$ is expressed as a linear combination $\sum_n u_n \hat{u}_n$ of finite-element basis functions $\hat{u}_n$, \eqref{gchan} becomes $\|\vec{o}^\dagger \vec{u}\|^2$ as in \eqref{Qchannel}, where $\vec{o}$ has components $o_n$ given by the linear functional
\begin{equation}
    o_n = o(\hat{u}_n) =\frac{1}{N_0}\int e_{y0} \hat{u}_n \mathrm{d}y\, .\label{eq:append_oi}
\end{equation}
Computationally, the assembly of $\vec{o}$ in finite-element software is equivalent to constructing a right-hand-side (source) vector~$\vec{b}$. 

The optimization {\rev{becomes}}:
\begin{eqnarray}
   g(\rho)&=&\max_{\rho} \left[\vec{u}(\rho)^\dagger\mat{B}(\rho)\vec{u}(\rho)\right]\, ,
   \, ,\nonumber\\
   \vec{u}(\rho) &=& \mat{A}(\rho)^{-\dagger}\vec{o}\,,\nonumber\\
   0&\leq&\rho\leq1\,. \label{eq:append_waveguide_objective}
\end{eqnarray}

By the adjoint method, for any $K$, we obtain the derivatives:
\begin{equation}
    \frac{\mathrm{d} g}{\mathrm{d}p}=\sum_{i=1}^K\left\{ \vec{u}_i^\dagger\frac{\mathrm{d}\mat{B}}{\mathrm{d}p}\vec{u}_i-2\Re\left[\vec{w}_i^\dagger\left(\frac{\mathrm{d}\mat{A}^\dagger}{\mathrm{d}p}\vec{u}_i\right)\right]\right\}\, ,\label{eq:adjointchan}
\end{equation}
where $\vec{w}_i$ solves $\mat{A}\vec{w}_i=\mat{B}\vec{u}_i$ and $\vec{u}_i$ solves the reciprocal problem $\mat{A}^\dagger\vec{u}_i = \vec{o}_i$ from \eqref{append_fewcha_obj}. {\rev{This derivative is also compared with the finite difference method and a difference of about $10^{-6}$ is observed. The analysis work flow is provided in \figref{flowchart2}.}}

\end{appendices}


\bibliography{reference}

\begin{thebibliography}{92}
\providecommand{\natexlab}[1]{#1}
\providecommand{\url}[1]{{#1}}
\providecommand{\urlprefix}{URL }
\providecommand{\doi}[1]{\url{https://doi.org/#1}}
\providecommand{\eprint}[2][]{\url{#2}}
 \bibcommenthead

\bibitem[{Aage and Egede~Johansen(2017)}]{Aage2017}
Aage N, Egede~Johansen V (2017) Topology optimization of microwave waveguide
  filters. International Journal for Numerical Methods in Engineering
  112(3):283--300. \doi{https://doi.org/10.1002/nme.5551}

\bibitem[{Agio and Cano(2013)}]{Purcell}
Agio M, Cano DM (2013) The {Purcell} factor of nanoresonators. Nat Phon
  7:674--675. \doi{10.1038/nphoton.2013.219}

\bibitem[{Azunre et~al(2019)Azunre, Jean, Rotschild, Bulovic, Johnson, and
  Baldo}]{AzunreJe19}
Azunre P, Jean J, Rotschild C, et~al (2019) Guaranteed global optimization of
  thin-film optical systems. New Journal of Physics 21:073,050

\bibitem[{Badia and Verdugo(2020)}]{gridap}
Badia S, Verdugo F (2020) Gridap: An extensible finite element toolbox in
  julia. J Open Source Softw 5(52):2520. \doi{10.21105/joss.02520}

\bibitem[{Bao et~al(2019)Bao, Cao, Lin, and Wyk}]{Bao2019}
Bao G, Cao Y, Lin J, et~al (2019) Computational optimal design of random rough
  surfaces in thin-film solar cells. Commun Comput Phys 25:1591--1612

\bibitem[{Basu et~al(2009)Basu, Zhang, and Fu}]{nearthermal}
Basu S, Zhang ZM, Fu CJ (2009) Review of near-field thermal radiation and its
  application to energy conversion. Int J Energy Res 33(13):1203--1232.
  \doi{10.1002/er.1607}

\bibitem[{Bayati et~al(2021)Bayati, Pestourie, Colburn, Lin, Johnson, and
  Majumdar}]{metalens}
Bayati E, Pestourie R, Colburn S, et~al (2021) Inverse designed extended depth
  of focus meta-optics for broadband imaging in the visible. Nanophotonics
  \doi{10.1515/nanoph-2021-0431}

\bibitem[{Bermel et~al(2010)Bermel, Ghebrebrhan, Chan, Yeng, Araghchini, Hamam,
  Marton, Jensen, Solja{\v{c}}i{\'{c}}, Joannopoulos, Johnson, and
  Celanovic}]{BermelGh10}
Bermel P, Ghebrebrhan M, Chan W, et~al (2010) Design and global optimization of
  high-efficiency thermophotovoltaic systems. Optics Express 18:A314--A334

\bibitem[{Bezanson et~al(2017)Bezanson, Edelman, Karpinski, and Shah}]{Julia}
Bezanson J, Edelman A, Karpinski S, et~al (2017) Julia: A fresh approach to
  numerical computing. SIAM {R}ev 59(1):65--98. \doi{10.1137/141000671}

\bibitem[{Brenny et~al(2014)Brenny, Coenen, and Polman}]{ICL}
Brenny BJM, Coenen T, Polman A (2014) Quantifying coherent and incoherent
  cathodoluminescence in semiconductors and metals. J Appl Phys
  115(24):244,307. \doi{10.1063/1.4885426}

\bibitem[{Capolino et~al(2007)Capolino, Jackson, Wilton, and Felsen}]{sumk}
Capolino F, Jackson DR, Wilton DR, et~al (2007) Comparison of methods for
  calculating the field excited by a dipole near a 2-d periodic material.
  {IEEE} Trans Antennas Propag 55(6):1644--1655. \doi{10.1109/TAP.2007.897348}

\bibitem[{Carey et~al(2008)Carey, Chen, Grigoropoulos, Kaviany, and
  Majumdar}]{farthermal}
Carey VP, Chen G, Grigoropoulos C, et~al (2008) A review of heat transfer
  physics. Nanoscale Microscale Thermophys Eng 12(1):1--60.
  \doi{10.1080/15567260801917520}

\bibitem[{Chew(2008)}]{Reciprocity}
Chew WC (2008) A new kook at reciprocity and energy conservation theorems in
  electromagnetics. {IEEE} Trans Antennas Propag 56(4):970--975.
  \doi{10.1109/TAP.2008.919189}

\bibitem[{Chong and Zak(2001)}]{OptBook}
Chong EKP, Zak SH (2001) An Introduction to Optimization, 2nd edn.
  Wiley-Interscience Publication

\bibitem[{Christiansen and Sigmund(2021)}]{RasmusTO}
Christiansen RE, Sigmund O (2021) Inverse design in photonics by topology
  optimization: tutorial. J Opt Soc Am B 38(2):496--509.
  \doi{10.1364/JOSAB.406048}

\bibitem[{Christiansen et~al(2020)Christiansen, Michon, Benzaouia, Sigmund, and
  Johnson}]{RamanTO}
Christiansen RE, Michon J, Benzaouia M, et~al (2020) Inverse design of
  nanoparticles for enhanced {R}aman scattering. Opt Express 28(4):4444--4462.
  \doi{10.1364/OE.28.004444}

\bibitem[{Davis(2006)}]{DavisBook}
Davis T (2006) Direct Methods for Sparse Linear Systems. SIAM

\bibitem[{Diaz and Sigmund(2010)}]{Diaz2010}
Diaz AR, Sigmund O (2010) A topology optimization method for design of negative
  permeability metamaterials. Struct Multidisc Optim 41:163--177.
  \doi{10.1007/s00158-009-0416-y}

\bibitem[{van Dijk et~al(2013)van Dijk, Maute, Langelaar, and
  Keulen}]{LevelSet}
van Dijk N, Maute K, Langelaar M, et~al (2013) Level-set methods for structural
  topology optimization: A review. Struct Multidiscipl Optim 48:437--472.
  \doi{10.1007/s00158-013-0912-y}

\bibitem[{Erchak et~al(2001)Erchak, Ripin, Fan, Rakich, Joannopoulos, Ippen,
  Petrich, and Kolodziejski}]{LED}
Erchak AA, Ripin DJ, Fan S, et~al (2001) Enhanced coupling to vertical
  radiation using a two-dimensional photonic crystal in a semiconductor
  light-emitting diode. Appl Phys Lett 78(5):563--565. \doi{10.1063/1.1342048}

\bibitem[{Ge et~al(2014)Ge, Kristensen, Young, and Hughes}]{Ge2014}
Ge RC, Kristensen PT, Young JF, et~al (2014) Quasinormal mode approach to
  modelling light-emission and propagation in nanoplasmonics. New Journal of
  Physics 16(11):113,048. \doi{10.1088/1367-2630/16/11/113048}

\bibitem[{Geuzaine and Remacle(2009)}]{gmsh}
Geuzaine C, Remacle JF (2009) Gmsh: A 3-d finite element mesh generator with
  built-in pre- and post-processing facilities. Int J Numer Methods Eng
  79(11):1309--1331. \doi{10.1002/nme.2579}

\bibitem[{Gibson(2021)}]{FMM}
Gibson WC (2021) The Method of Moments in Electromagnetics, 3rd edn. CRC Press,
  Boca Raton, FL

\bibitem[{Gong et~al(2021)Gong, Corrado, Mahbub, Shelden, and
  Munday}]{Casimir3}
Gong T, Corrado MR, Mahbub AR, et~al (2021) Recent progress in engineering the
  casimir effect--applications to nanophotonics, nanomechanics, and chemistry.
  Nanophotonics 10(1):523--536. \doi{10.1515/nanoph-2020-0425}

\bibitem[{Greffet et~al(2018)Greffet, Bouchon, Brucoli, and
  Marquier}]{Greffet2018}
Greffet JJ, Bouchon P, Brucoli G, et~al (2018) Light emission by nonequilibrium
  bodies: Local kirchhoff law. Phys Rev X 8:021,008

\bibitem[{Hackbusch(2015)}]{Hackbusch2015}
Hackbusch W (2015) Hierarchical Matrices: Algorithms and Analysis. Springer,
  Berlin

\bibitem[{Hammond et~al(2021)Hammond, Oskooi, Johnson, and
  Ralph}]{FoundryConstraint}
Hammond AM, Oskooi A, Johnson SG, et~al (2021) Photonic topology optimization
  with semiconductor-foundry design-rule constraints. Opt Express
  29:23,916--23,938. \doi{10.1364/OE.431188}

\bibitem[{Harrington(2001)}]{EquivalencePrinciple}
Harrington RF (2001) Time-Harmonic Electromagnetic Fields, 2nd edn. Wiley

\bibitem[{Hutchinson(1989)}]{Hutchinson}
Hutchinson M (1989) A stochastic estimator of the trace of the influence matrix
  for laplacian smoothing splines. Commun Stat B: Simul Comput
  18(3):1059--1076. \doi{10.1080/03610918908812806}

\bibitem[{Innes(2018)}]{Zygote}
Innes M (2018) Don't unroll adjoint: Differentiating ssa-form programs.
  Preprint at \url{https://arxiv.org/abs/1810.07951}

\bibitem[{Inui et~al(2012)Inui, Tanabe, and Onodera}]{Inui2012group}
Inui T, Tanabe Y, Onodera Y (2012) Group Theory and Its Applications in
  Physics. Springer, Berlin Heidelberg

\bibitem[{Janssen et~al(2010)Janssen, Wachters, and Urbach}]{Janssen2010}
Janssen OTA, Wachters AJH, Urbach HP (2010) Efficient optimization method for
  the light extraction from periodically modulated leds using reciprocity. Opt
  Express 18(24):24,522--24,535. \doi{10.1364/OE.18.024522}

\bibitem[{Jensen and Sigmund(2011)}]{TOReview}
Jensen J, Sigmund O (2011) Topology optimization for nano-photonics. Laser
  Photonics Rev 5(2):308--321. \doi{10.1002/lpor.201000014}

\bibitem[{Jin(2014)}]{FEMBook}
Jin J (2014) The Finite Element Method in Electromagnetics, 3rd edn.
  Wiley-{IEEE} Press

\bibitem[{Joannopoulos et~al(2008)Joannopoulos, Johnson, Winn, and
  Meade}]{PCBook}
Joannopoulos J, Johnson S, Winn J, et~al (2008) Photonic Crystals: Modeling the
  Flow of Light - Second Edition. Princeton University Press

\bibitem[{Johnson and Wichern(2018)}]{Correlation}
Johnson RA, Wichern DW (2018) Applied Multivariate Statistics, 6th edn. Pearson

\bibitem[{Johnson and Joannopoulos(2001)}]{JohnsonJo11}
Johnson S, Joannopoulos J (2001) Block-iterative frequency-domain methods for
  {Maxwell}'s equations in a planewave basis. Opt Express 8(3):173--190.
  \doi{10.1364/OE.8.000173}

\bibitem[{Johnson(2021)}]{NLopt}
Johnson SG (2021) The nlopt nonlinear-optimization package.
  \url{http://github.com/stevengj/nlopt}

\bibitem[{Johnson et~al(2005)Johnson, Povinelli, Solja{\v{c}}i{\'{c}}, Karalis,
  Jacobs, and Joannopoulos}]{JohnsonPo05}
Johnson SG, Povinelli ML, Solja{\v{c}}i{\'{c}} M, et~al (2005) Roughness losses
  and volume-current methods in photonic-crystal waveguides. Appl Phys B
  81(2--3):283--293. \doi{10.1007/s00340-005-1823-4}

\bibitem[{Kim(1986)}]{IncoherentSE}
Kim KJ (1986) An analysis of self-amplified spontaneous emission. Nucl Instrum
  250(1):396--403. \doi{10.1016/0168-9002(86)90916-2}

\bibitem[{Kita et~al(2018)Kita, Michon, Johnson, and Hu}]{KitaMi18}
Kita DM, Michon J, Johnson SG, et~al (2018) Are slot and sub-wavelength grating
  waveguides better than strip waveguides for sensing? Optica 5:1046--1054.
  \doi{10.1364/OPTICA.5.001046}

\bibitem[{Knyazev(2001)}]{LOBPCG}
Knyazev AV (2001) Toward the optimal preconditioned eigensolver: Locally
  optimal block preconditioned conjugate gradient method. SIAM J Sci Comput
  23(2):517--541. \doi{10.1137/S1064827500366124}

\bibitem[{Kokiopoulou et~al(2011)Kokiopoulou, Chen, and Saad}]{Saad2011}
Kokiopoulou E, Chen J, Saad Y (2011) Trace optimization and eigenproblems in
  dimension reduction methods. Numer Linear Algebra Appl 18(3):565--602.
  \doi{10.1002/nla.743}

\bibitem[{Kreutz-Delgado(2009)}]{CRCalculus}
Kreutz-Delgado K (2009) The complex gradient operator and the cr-calculus.
  Preprint at \url{https://arxiv.org/abs/0906.4835}

\bibitem[{Kuang and Miller(2020)}]{Kuang2020}
Kuang Z, Miller OD (2020) Computational bounds to light{\textendash}matter
  interactions via local conservation laws. Physical Review Letters 125(26)

\bibitem[{Lalanne et~al(2018)Lalanne, Yan, Vynck, Sauvan, and
  Hugonin}]{Lalanne2018}
Lalanne P, Yan W, Vynck K, et~al (2018) Light interaction with photonic and
  plasmonic resonances. Laser {\&} Photonics Reviews 12(5):1700,113

\bibitem[{Lanczos(1950)}]{Lanczos}
Lanczos C (1950) An iteration method for the solution of the eigenvalue problem
  of linear differential and integral operators. J Res Natl Bur Stand
  45(4):255--282. \doi{10.6028/JRES.045.026}

\bibitem[{Landau et~al(1980)Landau, Lif{\v{s}}ic, Lifshitz, P, Pitaevskii,
  Sykes, and Kearsley}]{FDT}
Landau L, Lif{\v{s}}ic E, Lifshitz E, et~al (1980) Statistical Physics: Theory
  of the Condensed State. Elsevier Science

\bibitem[{Lax(2013)}]{LinearAlgebra}
Lax P (2013) Linear Algebra and Its Applications. Wiley

\bibitem[{Lazarov and Sigmund(2011)}]{HelmholtzFilter}
Lazarov BS, Sigmund O (2011) Filters in topology optimization based on
  {Helmholtz}-type differential equations. Int J Numer Methods Eng
  86(6):765--781. \doi{10.1002/nme.3072}

\bibitem[{Li(2015)}]{rayleigh}
Li RC (2015) Rayleigh quotient based optimization methods for eigenvalue
  problems. In: Series in Contemporary Applied Mathematics. HEP, p 76--108

\bibitem[{Liang and Johnson(2013)}]{MaterialLossQ}
Liang X, Johnson SG (2013) Formulation for scalable optimization of
  microcavities via the frequency-averaged local density of states. Opt Express
  21(25):30,812--30,841. \doi{10.1364/OE.21.030812}

\bibitem[{Lin et~al(2022)Lin, Wang, and Hsu}]{lin2022full}
Lin HC, Wang Z, Hsu CW (2022) Full-wave solver for massively multi-channel
  optics using augmented partial factorization. arXiv preprint arXiv:220507887

\bibitem[{Luo et~al(2004)Luo, Narayanaswamy, Chen, and
  Joannopoulos}]{MonteCarlo3}
Luo C, Narayanaswamy A, Chen G, et~al (2004) Thermal radiation from photonic
  crystals: A direct calculation. Phys Rev Lett 93:213,905--213,908.
  \doi{10.1103/PhysRevLett.93.213905}

\bibitem[{Markovsky(2012)}]{LowRankApprox}
Markovsky I (2012) Low Rank Approximation. Springer-Verlag, London

\bibitem[{Miller et~al(2015)Miller, Johnson, and Rodriguez}]{MillerJo15}
Miller OD, Johnson SG, Rodriguez AW (2015) Shape-independent limits to
  near-field radiative heat transfer. Physical Review Letters 115:204,302

\bibitem[{Miller et~al(2016)Miller, Polimeridis, Reid, Hsu, DeLacy,
  Joannopoulos, Solja\v{c}i\'{c}, and Johnson}]{millerbound}
Miller OD, Polimeridis AG, Reid MTH, et~al (2016) Fundamental limits to optical
  response in absorptive systems. Opt Express 24(4):3329--3364.
  \doi{10.1364/OE.24.003329}

\bibitem[{Milonni(1976)}]{Milonni1976SemiclassicalAQ}
Milonni P (1976) Semiclassical and quantum-electrodynamical approaches in
  nonrelativistic radiation theory. Phys Rep 25:1--81.
  \doi{10.1016/0370-1573(76)90037-5}

\bibitem[{Molesky et~al(2018)Molesky, Lin, Piggott, Jin, Vucković, and
  Rodriguez}]{InverseDesign}
Molesky S, Lin Z, Piggott AY, et~al (2018) Inverse design in nanophotonics. Nat
  Photon 12(11):659--670. \doi{10.1038/s41566-018-0246-9}

\bibitem[{Molesky et~al(2020)Molesky, Venkataram, Jin, and
  Rodriguez}]{Rodriguez2020}
Molesky S, Venkataram PS, Jin W, et~al (2020) Fundamental limits to radiative
  heat transfer: Theory. Phys Rev B 101:035,408.
  \doi{10.1103/PhysRevB.101.035408}

\bibitem[{Mutapcic et~al(2009)Mutapcic, Boyd, Farjadpour, Johnson, and
  Avniel}]{MutapcicBo09}
Mutapcic A, Boyd S, Farjadpour A, et~al (2009) Robust design of slow-light
  tapers in periodic waveguides. Engineering Optimization 41:365--384

\bibitem[{Noda and Fujita(2009)}]{LEDReview}
Noda S, Fujita M (2009) Photonic crystal efficiency boost. Nature Photonics
  3:129--130. \doi{10.1038/nphoton.2009.15}

\bibitem[{Nussenzveig(1972)}]{HMBook}
Nussenzveig H (1972) Causality and Dispersion Relations. Academic Press, New
  York

\bibitem[{Oskooi and Johnson(2011)}]{PML}
Oskooi A, Johnson SG (2011) Distinguishing correct from incorrect {PML}
  proposals and a corrected unsplit {PML} for anisotropic, dispersive media. J
  Comput Phys 230:2369--2377. \doi{10.1016/j.jcp.2011.01.006}

\bibitem[{Oskooi and Johnson(2013)}]{OskooiJo13-sources}
Oskooi A, Johnson SG (2013) Electromagnetic wave source conditions. In: Taflove
  A, Oskooi A, Johnson SG (eds) Advances in {FDTD} Computational
  Electrodynamics: Photonics and Nanotechnology. Artech, Boston, chap~4, p
  65--100

\bibitem[{Pan et~al(2021)Pan, Christiansen, Michon, Hu, and Johnson}]{Raman3d}
Pan Y, Christiansen RE, Michon J, et~al (2021) Topology optimization of
  surface-enhanced {Raman} scattering substrates. Preprint at
  \url{https://arxiv.org/abs/2101.11352}

\bibitem[{Patra(2015)}]{NegaTemp}
Patra M (2015) On quantum optics of random media. PhD thesis, University of
  Leiden

\bibitem[{Payne and Lacey(1994)}]{Payne1994}
Payne FP, Lacey JPR (1994) A theoretical analysis of scattering loss from
  planar optical waveguides. Opt Quantum Electron 26:977--986.
  \doi{10.1007/BF00708339}

\bibitem[{Petersen and Pedersen(2012)}]{MatrixCookbook}
Petersen KB, Pedersen MS (2012) The Matrix Cookbook. Technical University of
  Denmark

\bibitem[{Pick et~al(2015)Pick, Cerjan, Liu, Rodriguez, Stone, Chong, and
  Johnson}]{PickCe15}
Pick A, Cerjan A, Liu D, et~al (2015) Ab-initio multimode linewidth theory for
  arbitrary inhomogeneous laser cavities. Phys Rev A 91:063,806.
  \doi{10.1103/PhysRevA.91.063806}

\bibitem[{Pilot et~al(2019)Pilot, Signorini, Durante, Orian, Bhamidipati, and
  Fabris}]{RamanReview}
Pilot R, Signorini R, Durante C, et~al (2019) A review on surface-enhanced
  {Raman} scattering. Biosensors 9(2):57. \doi{10.3390/bios9020057}

\bibitem[{Polimeridis et~al(2015)Polimeridis, Reid, Jin, Johnson, White, and
  Rodriguez}]{TraceFormula3}
Polimeridis AG, Reid MTH, Jin W, et~al (2015) Fluctuating volume-current
  formulation of electromagnetic fluctuations in inhomogeneous media:
  Incandescence and luminescence in arbitrary geometries. Phys Rev B
  92:134,202. \doi{10.1103/PhysRevB.92.134202}

\bibitem[{Reid et~al(2017)Reid, Miller, Polimeridis, Rodriguez, Tomlinson, and
  Johnson}]{TraceFormula4}
Reid MTH, Miller OD, Polimeridis AG, et~al (2017) Photon torpedoes and {Rytov}
  pinwheels: Integral-equation modeling of non-equilibrium fluctuation-induced
  forces and torques on nanoparticles. Preprint at
  \url{https://arxiv.org/abs/1708.01985}

\bibitem[{Reif(1965)}]{Kirchhoff}
Reif F (1965) Fundamentals of Statistical and Thermal Physics. McGraw-Hill
  Series in Fundamentals of Physics, New {Y}ork

\bibitem[{{Revels} et~al(2016){Revels}, {Lubin}, and
  {Papamarkou}}]{ForwardDiff}
{Revels} J, {Lubin} M, {Papamarkou} T (2016) Forward-mode automatic
  differentiation in {J}ulia. Preprint at
  \url{https://arxiv.org/abs/1607.07892}

\bibitem[{Rodriguez et~al(2011)Rodriguez, Ilic, Bermel, Celanovic,
  Joannopoulos, Solja{\v{c}}i{\'{c}}, and Johnson}]{MonteCarlo1}
Rodriguez AW, Ilic O, Bermel P, et~al (2011) Frequency-selective near-field
  radiative heat transfer between photonic crystal slabs: A computational
  approach for arbitrary geometries and materials. Phys Rev Lett 107:114,302.
  \doi{10.1103/PhysRevLett.107.114302}

\bibitem[{Rodriguez et~al(2013)Rodriguez, Reid, and Johnson}]{TraceFormula2}
Rodriguez AW, Reid MTH, Johnson SG (2013) Fluctuating surface-current
  formulation of radiative heat transfer: Theory and applications. Phys Rev B
  88:054,305. \doi{10.1103/PhysRevB.88.054305}

\bibitem[{Rogobete et~al(2003)Rogobete, Schniepp, Sandoghdar, and
  Henkel}]{SEinDielectricTheo}
Rogobete L, Schniepp H, Sandoghdar V, et~al (2003) Spontaneous emission in
  nanoscopic dielectric particles. Opt Lett 28(19):1736--1738.
  \doi{10.1364/OL.28.001736}

\bibitem[{Roques-Carmes et~al(2021)Roques-Carmes, Rivera, Ghorashi, Kooi, Yang,
  Lin, Beroz, Massuda, Sloan, Romeo, Yu, Joannopoulos, Kaminer, Johnson, and
  Solja{\v{c}}i{\'{c}}}]{Charles}
Roques-Carmes C, Rivera N, Ghorashi A, et~al (2021) A general framework for
  scintillation in nanophotonics. Preprint at
  \url{https://arxiv.org/abs/2110.11492}

\bibitem[{Schneider et~al(2019)Schneider, Santiago, Soltwisch, Hammerschmidt,
  Burger, and Rockstuhl}]{Schneider2019}
Schneider PI, Santiago XG, Soltwisch V, et~al (2019) Benchmarking five global
  optimization approaches for nano-optical shape optimization and parameter
  reconstruction. {ACS} Photonics 6(11):2726--2733

\bibitem[{Snyder and Love(1983)}]{OpticalWaveguide}
Snyder AW, Love JD (1983) Optical Waveguide Theory. Springer, USA

\bibitem[{Svanberg(2002)}]{svanberg2002class}
Svanberg K (2002) A class of globally convergent optimization methods based on
  conservative convex separable approximations. {SIAM} J Optim 12(2):555--573.
  \doi{10.1137/S1052623499362822}

\bibitem[{Tortorelli and Michaleris(1994)}]{DesignSensitivity}
Tortorelli DA, Michaleris P (1994) Design sensitivity analysis: Overview and
  review. Inverse Probl Eng 1(1):71--105. \doi{10.1080/174159794088027573}

\bibitem[{Trefethen and Bau(1997)}]{TrefethenBook}
Trefethen LN, Bau D (1997) Numerical Linear Algebra. SIAM

\bibitem[{Trefethen and Weideman(2014)}]{TrefethenExp}
Trefethen LN, Weideman JAC (2014) The exponentially convergent trapezoidal
  rule. SIAM Rev 56(3):385--458. \doi{10.1137/130932132}

\bibitem[{Ubaru et~al(2017)Ubaru, Chen, and Saad}]{FastEstimationTrace2017}
Ubaru S, Chen J, Saad Y (2017) Fast estimation of {$\mathrm{tr}(f(A))$} via
  stochastic {Lanczos} quadrature. SIAM J Matrix Anal Appl 38:1075--1099.
  \doi{10.1137/16M1104974}

\bibitem[{Wang et~al(2010)Wang, Lazarov, and Sigmund}]{FilterThreshold}
Wang F, Lazarov BS, Sigmund O (2010) On projection methods, convergence and
  robust formulations in topology optimization. Struct Multidiscipl Optim
  43(6):767--784. \doi{10.1007/s00158-010-0602-y}

\bibitem[{Wang et~al(2018)Wang, Christiansen, Yu, Mørk, and
  Sigmund}]{RasmusCavity}
Wang F, Christiansen RE, Yu Y, et~al (2018) Maximizing the quality factor to
  mode volume ratio for ultra-small photonic crystal cavities. Appl Phys Lett
  113(24):241,101. \doi{10.1063/1.5064468}

\bibitem[{Wolf(2007)}]{IncoherentImaging}
Wolf E (2007) Introduction to the Theory of Coherence and Polarization of
  Light. Cambridge University Press

\bibitem[{Yang et~al(2015)Yang, Wang, and Sun}]{WGM2D}
Yang S, Wang Y, Sun H (2015) Advances and prospects for whispering gallery mode
  microcavities. Adv Opt Mater 3(9):1136--1162. \doi{10.1002/adom.201500232}

\bibitem[{Yao et~al(2020)Yao, Benzaouia, Miller, and Johnson}]{LDOSOpt}
Yao W, Benzaouia M, Miller OD, et~al (2020) Approaching the upper limits of the
  local density of states via optimized metallic cavities. Opt Express
  28:24,185--24,197. \doi{10.1364/OE.397502}

\bibitem[{Yu et~al(2010)Yu, Raman, and Fan}]{SolarCell}
Yu Z, Raman A, Fan S (2010) Fundamental limit of nanophotonic light trapping in
  solar cells. {PNAS} 107(41):17,491--17,496. \doi{10.1073/pnas.1008296107}

\end{thebibliography}


\end{document}